\documentclass[11pt,dvips]{article}

\usepackage{jheppub}

\usepackage{slashbox,latexsym,amssymb,amsfonts,amsmath,slashed,bm}
\usepackage{graphicx}
\usepackage{mathrsfs}  
\newcommand{\f}[2]{\frac{#1}{#2}}

\newcommand{\la}{\langle}

\newcommand{\ra}{\rangle}

\renewcommand{\Re}{{\rm Re}\,}

\newcommand{\tr}{{\rm tr}\,}

\newcommand{\Oc}{{\cal O}}

\newcommand{\bred}{\bar{\beta}}

\title{Localisation in 2+1 dimensional SU(3) pure gauge theory at
  finite temperature}  

\author[a]{Matteo Giordano}

\affiliation[a]{
ELTE E\"otv\"os Lor\'and University, Institute for Theoretical
Physics, P\'azm\'any P.\ s.\ 1/A, H-1117, Budapest, Hungary
, and \\
MTA-ELTE Lend\"ulet Lattice Gauge Theory Research Group, P\'azm\'any
P.\ s.\ 1/A, H-1117, Budapest, Hungary}
\emailAdd{giordano@bodri.elte.hu}

\abstract{I study the localisation properties of low Dirac eigenmodes 
 in 2+1 dimensional SU(3) pure gauge theory, both in the
 low-temperature, confined and chirally-broken phase and in the
 high-temperature, deconfined and chirally-restored phase, by means of
 numerical lattice simulations. While these modes are delocalised at
 low temperature, they become localised at high temperature, up to a
 critical point in the Dirac spectrum where a BKT-type Anderson
 transition takes place. All results point to localisation appearing
 at the deconfinement temperature, and support previous expectations
 about the close relation between deconfinement, chiral symmetry breaking,
 and localisation. 
}

\keywords{Lattice Quantum Field Theory, Random Systems}

\begin{document}

\maketitle

\section{Introduction}
\label{sec:intro}

The close connection between deconfinement and chiral symmetry
restoration is an aspect of the finite-temperature transition in
QCD that is still poorly understood. As is well known, both the
confining and the chiral properties of QCD at vanishing chemical
potential change dramatically in the temperature range $T\simeq
145$--165 MeV: while in the low-temperature phase quarks and gluons 
are confined within hadrons and chiral symmetry is spontaneously
broken, in the high-temperature phase quarks and gluons are liberated
into a plasma, and chiral symmetry is restored. Of course, both
the spontaneous breaking and the restoration of chiral symmetry should
be understood here as approximate, since chiral symmetry is explicitly
(albeit softly) broken by the quark masses. Since the transition
is an analytic crossover~\cite{Borsanyi:2010bp}, there is no uniquely
defined critical temperature. However, defining a pseudocritical
temperature $T_\chi$ for the chiral transition as the position of the
peak of the chiral susceptibility, and a pseudocritical temperature
$T_{\rm dec}$ for the deconfinement transition as the position of the
peak of the quark entropy, one finds that they agree within errors,
$T_\chi=T_{\rm dec} = T_c\approx 155~{\rm
  MeV}$~\cite{Bazavov:2016uvm}. 

This relation between these two seemingly unrelated phenomena is not
unique to QCD, but it appears quite generally in a variety of gauge
theories. In particular, $T_\chi$ and $T_{\rm dec}$ coincide in
certain theories and models where the finite-temperature transition is
a genuine phase transition. Examples are provided by SU(3)
pure gauge theory in 3+1~\cite{Boyd:1996bx} and 2+1
dimensions~\cite{Damgaard:1998yv},\footnote{Although, strictly
  speaking, there is no chiral symmetry here since there are no
  fermions, one can nevertheless define a valence chiral condensate
  and study its behaviour in the limit of vanishing valence quark
  mass.} and $N_f=3$ QCD with unimproved rooted staggered fermions 
on $N_t=4$
lattices~\cite{Karsch:2001nf,deForcrand:2003vyj,deForcrand:2008vr}. 
Recent studies indicate that it is so also for the ${\cal N}=1$ SU(2)
super-Yang-Mills theory~\cite{Bergner:2019dim}. Another interesting
case is that of 3+1 dimensional SU(3) gauge theory with $N_f=2$
flavours of adjoint fermions~\cite{Karsch:1998qj} on the
lattice:\footnote{The continuum theory might lie inside the conformal
  window, but numerical studies have been inconclusive so
  far~\cite{DeGrand:2013uha}.} here two phase transitions are present,
a deconfining one at $T_{\rm dec}$ and a chiral-symmetry-restoring one
at $T_\chi$ with $T_{\rm dec}< T_\chi$. Nonetheless, at $T_{\rm dec}$
the chiral condensate jumps downwards, and a partial restoration of
chiral symmetry happens via a first-order phase transition. 

It is also well known that the fate of chiral symmetry is determined
by the spectrum of the Dirac operator near the origin. In fact, the
celebrated Banks-Casher relation~\cite{Banks:1979yr} establishes that
the chiral condensate in the chiral limit is proportional to the
spectral density of the Dirac operator near the origin. For finite but
small quark masses, an accumulation of eigenmodes near the origin is
still expected at low temperatures, leading to light pions and all the
other phenomenological consequences for QCD due to its being close to a 
theory with spontaneously broken symmetry. At high temperatures,
instead, the spectral density is expected to vanish near the origin,
reflecting the restoration of chiral symmetry in the massless case.
Given the close relation between confining and chiral properties of
the theory, it is natural to wonder if confinement is somehow
responsible for the accumulation of modes near the origin, and
analogously if deconfinement causes the depletion of this spectral
region. A similar question of course can be asked also for other gauge
theories.

Adding to the mistery, or possibly helping to solve it, a third
phenomenon has been observed to take place in QCD around the critical 
temperature, namely the localisation of the lowest modes of the Dirac
operator~\cite{GarciaGarcia:2005vj,GarciaGarcia:2006gr,Kovacs:2009zj,
  Kovacs:2010wx,Kovacs:2012zq,Cossu:2016scb,Holicki:2018sms}. 
Numerical studies on the lattice have shown that while in the
low-temperature phase all the Dirac modes are extended throughout the
whole system, above $T_c$ the lowest modes get
localised~\cite{GarciaGarcia:2005vj,GarciaGarcia:2006gr,Kovacs:2009zj, 
  Kovacs:2010wx,Kovacs:2012zq,Cossu:2016scb,Holicki:2018sms}
on the scale of the inverse temperature~\cite{Kovacs:2012zq}. More
precisely, modes are localised up to a temperature-dependent critical
point in the spectrum, $\lambda_c=\lambda_c(T)$.\footnote{Here and in
  the following, with a slight abuse of terminology, I will call
  ``eigenvalue'' what is really the imaginary part of the eigenvalue.}
At $\lambda_c$, a second-order phase transition takes place in the
spectrum~\cite{Giordano:2013taa}, and modes become delocalised. This
type of transitions is well known in the condensed-matter literature
as {\it Anderson transitions}~\cite{lee1985disordered,Evers:2008zz},
and the corresponding critical point is known as {\it mobility
  edge}. The mobility edge, $\lambda_c$, vanishes at a temperature
compatible with $T_c$~\cite{Kovacs:2012zq}, suggesting that
localisation of the low Dirac modes is closely related to
deconfinement and chiral restoration. This is further supported by a
similar coincidence of the three phenomena in other theories and
models, like SU(3) pure gauge theory in 3+1
dimensions~\cite{Kovacs:2017uiz}, the $N_f=3$ unimproved staggered
fermion model mentioned above~\cite{Giordano:2016nuu}, and also in a
toy model for QCD~\cite{Giordano:2016vhx}, devised in
Ref.~\cite{Giordano:2016cjs} precisely to study the issue of
localisation. 

A qualitative understanding of the relation between deconfinement and
localisation is provided by what in this paper will be referred to as
the ``sea/islands'' picture of
localisation~\cite{Bruckmann:2011cc,Giordano:2015vla}. The idea is
that the local Polyakov lines provide a sort of local potential for
the Dirac modes via the effective boundary condition that they impose
on the eigenmodes. In the high-temperature phase, this looks like a
``sea'' corresponding to the Polyakov lines ordered along the
identity, in which ``islands'' corresponding to non-aligned Polyakov
lines appear. Such islands are ``energetically'' favourable, and thus 
provide convenient places where the eigenmodes can localise. This
picture is made more precise by exactly recasting the staggered Dirac
operator as the Hamiltonian of a set of coupled three-dimensional
Anderson models~\cite{Giordano:2016cjs}, with the phases of the local
Polyakov lines acting as the source of a random on-site potential, and
with the strength of the coupling among the different Anderson models 
decreasing as the system gets ordered. Notice that the dimensionality
of the relevant Anderson models matches the spatial dimension of the
gauge theory. The ``Dirac-Anderson'' form of the staggered operator
suggests that the strength of the coupling plays an important role for
the fate of localisation, as well as for the spontaneous breaking of
chiral symmetry: as the Anderson models decouple, it becomes more
difficult for modes to accumulate around the origin. These ideas are
supported by the results of a detailed study on variations of a toy
model for QCD~\cite{Giordano:2016cjs}.   

An interesting aspect of the sea/islands picture and of the
Dirac-Anderson approach to localisation and chiral symmetry
restoration is that they depend only marginally on the dimensionality
of the system, or the gauge group, or the type of fermions present in
the theory: all that is required for the argument to apply is
essentially the existence of an ordered phase with Polyakov lines
aligning to the identity, and the possibility for modes to localise in
the relevant Anderson model. A non trivial test of these ideas can then
be performed by studying gauge theories in different dimensions, with
different gauge groups, and different fermion representations. In this
paper we try the first possibility, studying SU(3) pure gauge theory
in 2+1 dimensions on the lattice using staggered fermions. As already
mentioned above, this theory is known to display a deconfining and
chiral-symmetry-restoring second-order phase transition at finite
temperature.   

The choice of dimensionality is somewhat peculiar, both for chiral
symmetry and localisation. As a matter of fact, chiral symmetry as
usually defined does not even exist in odd dimensions. However, 
in three dimensions for an even number of flavours, $N_f$, it is
possible to reorganise the $N_f$ two-component spinors into $N_f/2$
four-component spinors, and in the massless case the continuum Dirac
action is invariant under a $\text{U}(N_f)$ flavour symmetry group
with two ``chiral''
subgroups~\cite{Pisarski:1984dj,Burden:1986by}.\footnote{This 
  construction generalises to any odd dimension $D$, reorganising the
  $N_f$ $2^{\f{D-1}{2}}$-component spinors into $\f{N_f}{2}$
  $2^{\f{D+1}{2}}$-component spinors~\cite{Burden:1986by}.} While these 
are explicitly broken by a mass term, in the massless case one 
can meaningfully ask if they are spontaneously broken due to the
formation of a quark-antiquark condensate, which breaks the flavour
symmetry down to $\text{U}(1)\times\text{U}(1)\times
\text{SU}(N_f/2)\times \text{SU}(N_f/2)$~\cite{Pisarski:1984dj}. The
Banks-Casher relation then ties the spontaneous breaking of this
symmetry to the accumulation of Dirac eigenmodes around
zero. Concerning localisation, in two dimensions the existence of an
Anderson transition from localised to delocalised modes in a
disordered system depends on the details of the model (see, e.g.,
Ref.~\cite{Evers:2008zz}). Using one flavour of staggered fermions,
one has effectively $N_f=2$ in the continuum due to the doubling
phenomenon, so that chiral symmetry (in the above sense) can be
defined; using SU(3) as gauge group, the symmetry class in the
classification of Random Matrix Theory is the unitary one, for which
Anderson transitions in two spatial dimensions are known to
exist~\cite{xie1998kosterlitz}.  

As there is no obstruction to the sea/islands picture to work, then it 
should just work, leading to the same situation encountered in 3+1
dimensions: at low temperature, chiral symmetry should be
spontaneously broken by the accumulation of delocalised Dirac
eigenmodes near the origin; at high temperature, the spectral density
should vanish near the origin, leading to chiral symmetry restoration,
and modes should be localised up to a mobility edge somewhere in the
spectrum. It is already known that chiral symmetry is restored at
deconfinement~\cite{Damgaard:1998yv}. It is the purpose of this work
to verify that the localisation properties of the low modes change
there as well. 

The plan of this paper is the following. In Section \ref{sec:loclgt} I 
briefly review localisation in disordered systems, with a special
focus on lattice gauge theories in 3+1 dimensions. In Section
\ref{sec:su3_d} I describe in some detail the specific model under
consideration, and expectations about its behaviour. In Section
\ref{sec:numres} I show numerical results and their
analysis. Conclusions and perspectives on future investigations are
discussed in Section \ref{sec:concl}.

\section{Localisation in lattice gauge theories}
\label{sec:loclgt}

Localisation is a well known phenomenon in condensed matter
physics. It has long been known, since the seminal work of
Anderson~\cite{Anderson:1958vr}, that the addition of a random on-site
potential to the usual tight-binding Hamiltonian causes the
localisation of the energy states at the band edge, beyond a critical
energy $E_c$ called mobility edge, while states in the band centre
remain extended (see Ref.~\cite{lee1985disordered} for a review). Such
``disordered Hamiltonians'' aim at describing metals with impurities,
and in this context the width $W$ of the distribution 
of the on-site potential is a measure of the amount of impurities in
the system. As the disorder parameter $W$ is increased, the mobility
edge moves towards the band centre, and all modes become localised
beyond some critical disorder, turning the metal into an
insulator. In three (and higher) dimensions the transition between
localised and delocalised modes is a second-order quantum phase
transition known as Anderson transition (see Ref.~\cite{Evers:2008zz}
for a review), with a divergent correlation length $\xi \sim
|E-E_c|^{-\nu}$ characterised by the critical exponent $\nu$. 

Localisation in gauge theories was initially studied at zero
temperature in investigations of the topological structure of the QCD 
vacuum (see the review Ref.~\cite{deForcrand:2006my} and references
therein). The idea that the finite-temperature transition of QCD could
be related to localisation of the low Dirac modes dates back to 
Refs.~\cite{Diakonov:1985eg,Diakonov:1995ea,Halasz:1995vd}. The first
numerical results supporting this idea appeared in
Ref.~\cite{GarciaGarcia:2005vj}, coming from the effective 
description of QCD via an Instanton Liquid Model, and in
Ref.~\cite{GarciaGarcia:2006gr}, coming from numerical simulations of
QCD on a lattice, both in the quenched approximation and with 2+1
flavours of staggered fermions. Further evidence of localisation of
the low Dirac modes in the high-temperature phase of QCD was provided
by the absence of correlations in the low-lying spectrum of the
overlap operator~\cite{Kovacs:2009zj}, typical of localised modes. 

A detailed study of localisation in lattice QCD was undertaken in
Ref.~\cite{Kovacs:2012zq}, using 2+1 flavours of 2-stout improved
rooted staggered fermions. There, it was shown that above $T_c$ the
eigenmodes of the staggered Dirac operator are localised, for
(imaginary part of the) eigenvalue up to a temperature-dependent
mobility edge, $\lambda_c(T)$, beyond which modes become extended.
The extrapolation of $\lambda_c(T)$ vanishes at a temperature
compatible with $T_c$, in agreement with the absence of localised
modes in the low-temperature phase. Localisation was shown to survive
the continuum limit, indicating that it is not a lattice
artefact. This is also supported by the fact that localised modes have
been found with other fermion discretisations, namely with domain-wall 
fermions~\cite{Cossu:2016scb} and with overlap fermions on
twisted-mass Wilson fermion backgrounds~\cite{Holicki:2018sms}.

Localisation in the high-temperature, deconfined phase has been
observed also in other gauge theories (see
Ref.~\cite{Giordano:2018iei} for a recent review), namely 
SU(2)~\cite{Kovacs:2010wx} and SU(3) pure gauge  
theory~\cite{Kovacs:2017uiz}, and SU(3) with $N_f=3$ flavours of
unimproved rooted staggered fermions on $N_t=4$
lattices~\cite{Giordano:2016nuu}. In these models the
finite-temperature transition is a genuine phase transition, which
provides a clean-cut setting for investigating the possible
coincidence of deconfinement, chiral-symmetry restoration and
localisation of the low Dirac modes. While in the first case a
detailed study of the temperature dependence of the mobility edge is
missing, in the other two cases it was indeed found that the low Dirac
modes start localising precisely at the critical temperature where
deconfinement and chiral-symmetry restoration take place.
A similar result was found in a toy model for 
QCD~\cite{Giordano:2016vhx}, where the Polyakov-line dynamics is 
mimicked by a spin model: as the spins get ordered, the low Dirac
modes localise and their density near the origin drops to zero.

The simplest way to identify localised modes is by studying their
so-called {\it participation ratio} (PR). Given a normalised
eigenmode of a lattice  Dirac operator, $\Psi(n)$, one defines the
inverse participation ratio (IPR) as
\begin{equation}
  \label{eq:IPR}
  {\rm IPR} = \sum_n |\Psi(n)^\dag \Psi(n)|^2\,,
\end{equation}
where $\Psi(n)^\dag \Psi(n)$ denotes the scalar product in colour and
(possibly) Dirac space, and the sum is over the lattice sites
$n$.\footnote{\label{foot:genIPR} We mention in passing the
  generalised IPRs, defined as 
${\rm IPR}_q \equiv \sum_n |\Psi(n)^\dag \Psi(n)|^{q}$. Clearly ${\rm
  IPR} = {\rm IPR}_2$.}
The PR is just the inverse of the IPR divided by the lattice size,
\begin{equation}
  \label{eq:PR}
  {\rm PR} = \f{{\rm IPR}^{-1}}{N_t V}\,, \qquad V= N_s^d\,,
\end{equation}
where we have assumed that the lattice is a $(d+1)$-dimensional
hypercube of spatial extension $N_s$ and temporal extension $N_t$ in
lattice units. In all the models discussed above $d=3$, while in this 
work I will consider $d=2$. For a fully delocalised mode, 
$\Psi(n)^\dag \Psi(n)\sim 1/V$, so that the ${\rm PR}$ remains
constant in the large volume limit.\footnote{With ``volume'' I will
  refer to the spatial volume $V=N_s^d$, unless it is explicitly
  stated otherwise.} For a mode localised in a spatial region of fixed
size $v$ one finds instead ${\rm PR} \sim v/V\to 0$ in the
large-volume limit. 

A useful observation is that in a random matrix model the localisation
properties of the eigenmodes and the statistical properties of the
corresponding eigenvalues are closely
related~\cite{altshuler1986repulsion}. For localised modes the 
eigenvalues fluctuate independently, following Poisson statistics,
while for extended modes the eigenvalue statistics are those of the
appropriate Gaussian ensemble of Random Matrix Theory (RMT; see
Ref.~\cite{mehta2004random} for a general introduction, and
Refs.~\cite{Guhr:1997ve,Verbaarschot:2000dy} for the application of
RMT to QCD). This is made evident after unfolding the spectrum, i.e.,
after mapping  $\lambda_i \to x_i=\int^{\lambda_i}d\lambda'
\,\rho(\lambda')$, where $\rho(\lambda)\equiv \la
\sum_i\delta(\lambda-\lambda_i)\ra$ is the spectral density and
$\la\ldots\ra$ denotes averaging over the random matrix ensemble.
This mapping makes the spectral density identically 1 throughout the 
spectrum. It is well known that for dense matrices the (bulk)
statistical properties of the unfolded spectrum are universal and
uniform throughout the spectrum, i.e., do not depend on the details of
the random matrix model under study, but only on the symmetry class of
the ensemble~\cite{mehta2004random}. The main classes are the
orthogonal, unitary, and symplectic classes.\footnote{The
  classification in symmetry classes of RMT ensembles is actually
  richer~\cite{Zirnbauer:2010gg}, but these ensembles suffice for our
  purposes.} Universal analytic results can therefore be obtained
using the so-called Gaussian ensembles. In particular, the probability 
distribution of consecutive unfolded level spacings $s_i\equiv
x_{i+1}-x_i$ is known, and very accurately approximated by the
so-called {\it Wigner surmise}, which for the unitary class reads 
\begin{equation}
  \label{eq:WS}
P_{\rm RMT}(s) = \f{32}{\pi^2}s^2 
 e^{-\f{4}{\pi}s^2}\,.
\end{equation}
In contrast, for eigenvalues that fluctuate independently the unfolded
level spacing distribution is the exponential function, appropriate
for Poisson statistics,
\begin{equation}
  \label{eq:Pois}
P_{\rm Poisson}(s) = e^{-s}\,.
\end{equation}
The lattice Dirac operator in a gauge-field background is in practice
a sparse random matrix, with fluctuations provided by the gauge links
and ensemble averaging corresponding to integration over gauge 
fields with the appropriate measure. For ensembles of sparse matrices
the statistical properties can depend on the spectral region under
consideration, and so it is customary to compute the spectral
statistics locally in the spectrum, i.e., restricting to small
spectral intervals around the chosen point. Looking at Dirac spectra
in the high-temperature phase of QCD one sees indeed a transition from
Poisson to RMT statistics. By a finite-size scaling study of
statistical spectral observables it is possible to determine precisely
the location of the mobility edge and the critical exponent
$\nu$~\cite{Shklovskii:1993zz}.  This has been used to show that the
localisation/delocalisation transition in the Dirac spectrum in
high-temperature lattice QCD with staggered fermions is a genuine
second-order phase transition in the same universality
class~\cite{Giordano:2013taa} as the three-dimensional unitary
Anderson model~\cite{slevin1999corrections}. Further support to this
conclusion came from a study of the multifractal properties of
eigenmodes at the mobility edge~\cite{Ujfalusi:2015nha}. This matching
can be easily understood in the light of the sea/islands 
picture~\cite{Bruckmann:2011cc,Giordano:2015vla} and of the
Dirac-Anderson approach~\cite{Giordano:2016cjs}, since the spatial
fluctuations of the Polyakov lines precisely provide the kind of
three-dimensional on-site disorder present in the Anderson model,
while the unitary symmetry class is the one to which the staggered
Dirac operator belongs.

\section{Lattice SU(3) pure gauge theory in 2+1 dimensions}
\label{sec:su3_d}

Finite-temperature SU(3) pure gauge theory in 2+1 dimensions has been
studied on the lattice in several
papers~\cite{Gross:1984pq,Christensen:1991rx,Christensen:1992is,Engels:1996dz,
Liddle:2008kk,Bialas:2008rk,Bialas:2009pt,Bialas:2012qz}. This
theory shows a deconfining second-order phase transition at finite
temperature, in the same universality class as the two-dimensional
3-colour Potts model. These works used a hypercubic lattice and the
Wilson action, which up to an irrelevant additive constant reads 
\begin{equation}
  \label{eq:wilson_act}
  S[U] = -\f{\beta}{3}\sum_n\sum_{\mu<\nu} \Re\tr U_{\mu\nu}(n) =
  -\bred\sum_n\sum_{\mu<\nu} \Re\tr U_{\mu\nu}(n) \,,
\end{equation}
where  $\bred=\beta/3$, $U_{\mu\nu}(n)$ is the usual plaquette variable,
\begin{equation}
  \label{eq:wilson_act2}
  U_{\mu\nu}(n) = U_\mu(n)U_\nu(n+\hat{\mu})U_\mu(n+\hat{\nu})^\dag
  U_\nu(n)^\dag\,, 
\end{equation}
$\mu,\nu=1,2,3$ are the lattice directions, $U_\mu(n)$ are SU(3)
gauge links living on the lattice edges $(n,n+\hat{\mu})$, and
$\hat{\mu}$ denotes the unit vector in direction $\mu$. Periodic 
boundary conditions both in the temporal and in the spatial directions
are imposed. In 2+1 dimensions the gauge coupling $g$ has dimensions
of ${\rm [mass]}^{\f{1}{2}}$, and the lattice coupling $\beta$ is related to
$g$ and the lattice spacing as $\beta=6/(g^2a)$.\footnote{For the sake
of simplicity we ignore scaling violations.} The partition function is
\begin{equation}
  \label{eq:wilson_act3}
  Z = \int \mathscr{D}U\,e^{-S[U]}\,, \qquad \mathscr{D} U=\prod_{n,\mu}
  dU_\mu(n)\,,
\end{equation}
with $dU_\mu(n)$ the SU(3) Haar measure. The critical temperature was
determined precisely in Ref.~\cite{Liddle:2008kk}, and for lattices of
temporal extension $N_t=4$ it corresponds to the lattice gauge coupling
$\bred_c = 4.9057(57)$. 

Chiral symmetry breaking, as discussed in the Introduction, was first
studied in Ref.~\cite{Damgaard:1998yv} with staggered fermions in the
quenched approximation, so studying the Dirac operator on pure-gauge
theory backgrounds. The staggered Dirac operator reads
\begin{equation}
  \label{eq:stag}
  D^{\rm stag}_{n,n'} = \f{1}{2}\sum_\mu
  \eta_\mu(n)\left(U_\mu(n)\delta_{n+\hat{\mu},n'} 
-U_\mu(n-\hat{\mu})^\dag\delta_{n-\hat{\mu},n'}\right)\,,\qquad
\eta_\mu(n) = (-1)^{\sum_{\nu<\mu}n_\mu}\,,
\end{equation}
with periodic boundary conditions in the spatial directions and
antiperiodic boundary conditions in the temporal direction, 
and in the continuum limit it describes $N_f=2$ degenerate species of
fermions. The staggered operator $D^{\rm stag}$ has purely imaginary
eigenvalues, $i\lambda$, and since its spectrum is symmetric about
zero it is enough to consider only $\lambda\ge 0$. Fermions break the
centre symmetry of the pure-gauge theory, selecting the vacuum with
trivial Polyakov lines, but when working in the quenched approximation
this has to be done by hand, for example by multiplying all the
temporal links in the last time slice by the appropriate centre
element. This was the approach adopted in Ref.~\cite{Damgaard:1998yv},
which we will also use when needed. While the full chiral subgroup of
the $\text{U}(N_f)$ flavour symmetry discussed in the Introduction is
explicitly broken at finite lattice spacing, a remnant
$\text{U}(1)\times \text{U}(1)$ chiral symmetry still
survives~\cite{Burden:1986by}, and can be spontaneously broken by the
formation of a quark-antiquark condensate. The authors of
Ref.~\cite{Damgaard:1998yv} observed that while in the low $\beta$
(low temperature), confined phase a finite chiral condensate was found
in the limit of vanishing valence quark mass, this vanished in the
high $\beta$, deconfined phase, and that the chiral transition
coincided with the deconfinement transition.  

The purpose of this work is to study the localisation properties of
the low Dirac modes in the two phases of the theory, both by measuring
the PR of the eigenmodes, Eq.~\eqref{eq:PR}, and by studying the
statistical properties of the unfolded spectrum. After obtaining the
Dirac spectra for an ensemble of gauge configurations, unfolding is
done by sorting all the eigenvalues (for a given coupling and lattice
size) by increasing size, and replacing them by their rank divided by
the number of configurations. This automatically makes the spectral
density equal to 1 throughout the spectrum. I use two quantities
extracted from the unfolded level spacing distribution measured
locally in the spectrum, $P_\lambda(s)$. One of them is the integrated
distribution $I_\lambda$,  
\begin{equation}
  \label{eq:Ilambda}
  I_\lambda = \int_0^{s_0}ds\,P_\lambda(s)\,,
\end{equation}
where $s_0\simeq 0.508$ is conveniently chosen to be 
the crossing point of the unitary Wigner surmise, Eq.~\eqref{eq:WS},
and the exponential function, Eq.~\eqref{eq:Pois}, i.e., to a very good
approximation, the crossing point of the unfolded level spacing 
distributions for RMT and Poisson statistics. This maximises the 
difference between the two types of behaviour. The analytic
predictions for $I_\lambda$ in the two cases are $I_{\rm RMT} \simeq
0.117$ and $I_{\rm Poisson} \simeq 0.398$. The other quantity is the
second moment of the distribution, $\la s^2\ra_\lambda$,
\begin{equation}
  \label{eq:Ilambda2}
\la s^2\ra_\lambda = \int_0^{\infty} ds\,P_\lambda(s)s^2\,,
\end{equation}
which takes the values $\la s^2\ra_{\rm RMT} \simeq \f{3\pi}{8}$ and $\la
s^2\ra_{\rm Poisson} = 2$ for the two types of statistics.

In the thermodynamic limit, the spectral density and the local average 
level spacing $\la \Delta\lambda\ra_\lambda \equiv\la
\lambda_{i+1}-\lambda_i\ra_\lambda$ are related as $\rho(\lambda)\la 
\Delta\lambda\ra_\lambda =1$. The same relation holds for the unfolded
spectrum, and so, since the unfolded spectral density equals 1, one
has $\la s \ra_\lambda \equiv \la \Delta x\ra_\lambda=1$ in infinite
volume. This might however fail in a finite volume, where the local
averaging is necessarily done in a small but finite spectral
interval,\footnote{In this case, when computing the average of
  $\lambda_{i+1}-\lambda_i$ we ask that $\lambda_i$ be inside the
  interval, while $\lambda_{i+1}$ can be outside.} in 
regions where the spectral density is small and rapidly varying.
Indeed, it is easy to see that the average level spacing in an
interval of size $\Delta$ is given by the difference between the 
smallest eigenvalue $\lambda'$ right above the end of the interval and
the smallest eigenvalue $\lambda$ inside the interval, divided by the
number $N$ of eigenvalues inside the interval. The spectral density
associated to the given interval is simply $N/\Delta$. If the spectral
density is increasing, then one typically finds $\lambda'-\lambda <
\Delta$, since eigenvalues get closer going up in the spectrum, and so 
$(N/\Delta)\cdot [(\lambda'-\lambda)/N] < 1$, from which we are led to
expect $\rho(\lambda)\la \Delta\lambda\ra_\lambda <1$, and so $\la s
\ra_\lambda < 1$. This is what happens near the origin in the high
temperature phase, where $\rho(\lambda)\sim \lambda^\alpha $ for some
positive power $\alpha$: the small spectral density requires the use
of relatively large intervals over which the spectral density increases
non-negligibly. The value of $\la s \ra_\lambda$ can then be used to
assess the reliability of estimates of spectral statistics based on
unfolding, and only those spectral regions where $\la s \ra_\lambda
\simeq 1$ should be considered in further analyses.  

From the point of view of the Dirac-Anderson approach, the model at
hand should behave like a two-dimensional unitary Anderson model with
on-site (diagonal) disorder.\footnote{Off-diagonal disorder is present
as well, but does not play an important role in localising the
eigenmodes~\cite{economou1977localization,weaire1977numerical}.} This
model has been studied in Ref.~\cite{xie1998kosterlitz}, where it was
shown that it displays an Anderson transition of
Berezinski\u{\i}-Kosterlitz-Thouless (BKT) 
type~\cite{Berezinsky:1970fr,Berezinsky:1972fr,Kosterlitz:1973xp}, 
with an exponentially divergent localisation length. We expect to find
the same behaviour in our model in the high-$\beta$ phase, with a
mobility edge $\lambda_c$ separating localised (low) and delocalised
(high) Dirac modes, where the localisation length should diverge in
the infinite-volume limit as 
\begin{equation}
  \label{eq:xi_BKT}
  \xi \sim \exp\left\{\f{\alpha}{\sqrt{\lambda_c-\lambda}}\right\}\,,
\end{equation}
for some constant $\alpha$, as one approaches $\lambda_c$ from 
the localised side, $\lambda<\lambda_c$. Points on the delocalised
side, $\lambda>\lambda_c$, are all critical in a disorder-driven BKT
transition, and so there the statistical properties of the spectrum
should be independent of the volume (see
Ref.~\cite{barber1983finite}). To verify if this is the case, I did a 
finite-size scaling analysis of the spectral observables
$\Oc_\lambda=I_\lambda,\la s^2\ra_\lambda$, making the scaling hypothesis  
\begin{equation}
  \label{eq:BKT_scaling}
  \Oc_\lambda(L) = F\left((\lambda-\lambda_c)(\log
    L)^{\f{1}{\nu}}\right)\,, 
\end{equation}
with $F$ some analytic function and $L$ the linear size of the
system. The one-parameter scaling hypothesis for
localisation~\cite{locgangoffour} is usually motivated by assuming 
that the behaviour of the system near the transition in a 
finite volume is determined only by the ratio $\xi(\lambda)/L$. However,
since we are working on a lattice, $\xi$ and $L$ can be made
dimensionless dividing them by the lattice spacing, and so one can use
instead a function $F(r)$ of ratios of the more general form
$r=h(\xi(\lambda))/h(L)$. Indeed, since near the
transition and for large volumes only the leading behaviour of $h$
matters, $r$ is invariant under the rescaling $\xi \to b\xi$, $L\to
bL$, as it should. On the other hand, analyticity in a finite
volume constrains the form of $h$, which in the case at hand must be
of the form $h(\xi) = 1/(\log\xi)^{\f{1}{\nu}}$, where $\nu=\f{1}{2}$ if
Eq.~\eqref{eq:xi_BKT} holds.

\section{Numerical results}
\label{sec:numres}

In this Section I report on the results of a numerical investigation
of Dirac spectra in finite-temperature SU(3) pure-gauge theory in 2+1
dimensions on the lattice. The gauge backgrounds were obtained via
Monte Carlo simulations of the partition function
Eq.~\eqref{eq:wilson_act3}, using a hypercubic lattice of fixed
temporal extension $N_t=4$ and spatial extension
$N_s=32,40,48,56,64,72$ in lattice units, for several values of 
$\bred$, both in the confined and the deconfined phases. Right
above the deconfinement transition, and for the lattice sizes used
here, a system initialised in the trivial Polyakov-loop sector is
still able to switch to one of the complex sectors, and to temporarily
tunnel to the confined phase, even though this becomes less and less
likely as the volume is increased. For
$\bred=5.05,\,5.10,\,5.15,\,5.20$ I then always analysed the 
gauge configuration obtained by rotating the centre sector to the
trivial one, as explained above in Section \ref{sec:su3_d}. For  
$\bred=5.25,\,5.50,\,5.75,\,6.00,\,6.25,\,6.50$ this was not
necessary, as a system initialised in the trivial Polyakov-loop sector 
never left it. Of course, no sector switching was used for
$\bred=2.50,\,3.00,\,4.00$ in the confined phase. The first few
eigenvalues of the staggered Dirac operator were obtained by means of
the ARPACK routine~\cite{lehoucq1998arpack}.  In the following I
denote with $\lambda$ the (imaginary part of the) eigenvalues of the
staggered Dirac operator in lattice units. Monte Carlo simulations
were performed with standard
heatbath~\cite{Creutz:1980zw,Cabibbo:1982zn} and
overrelaxation~\cite{Brown:1987rra,Creutz:1987xi} algorithms. Details  
about statistics and the number of computed eigenvalues for each
volume can be found in Table \ref{tab:details}. All statistical errors
were obtained through a jackknife analysis with 100 samples. All fits
in this work were done using the MINUIT routine~\cite{James:1975dr},
establishing the accuracy of the errors with the MINOS
subroutine. Whenever the difference is negligible, the symmetric
parabolic error is used instead.

\subsection{Participation ratio}
\label{sec:numpr}

\begin{figure}[t]
  \centering
  \includegraphics[width=0.75\textwidth]{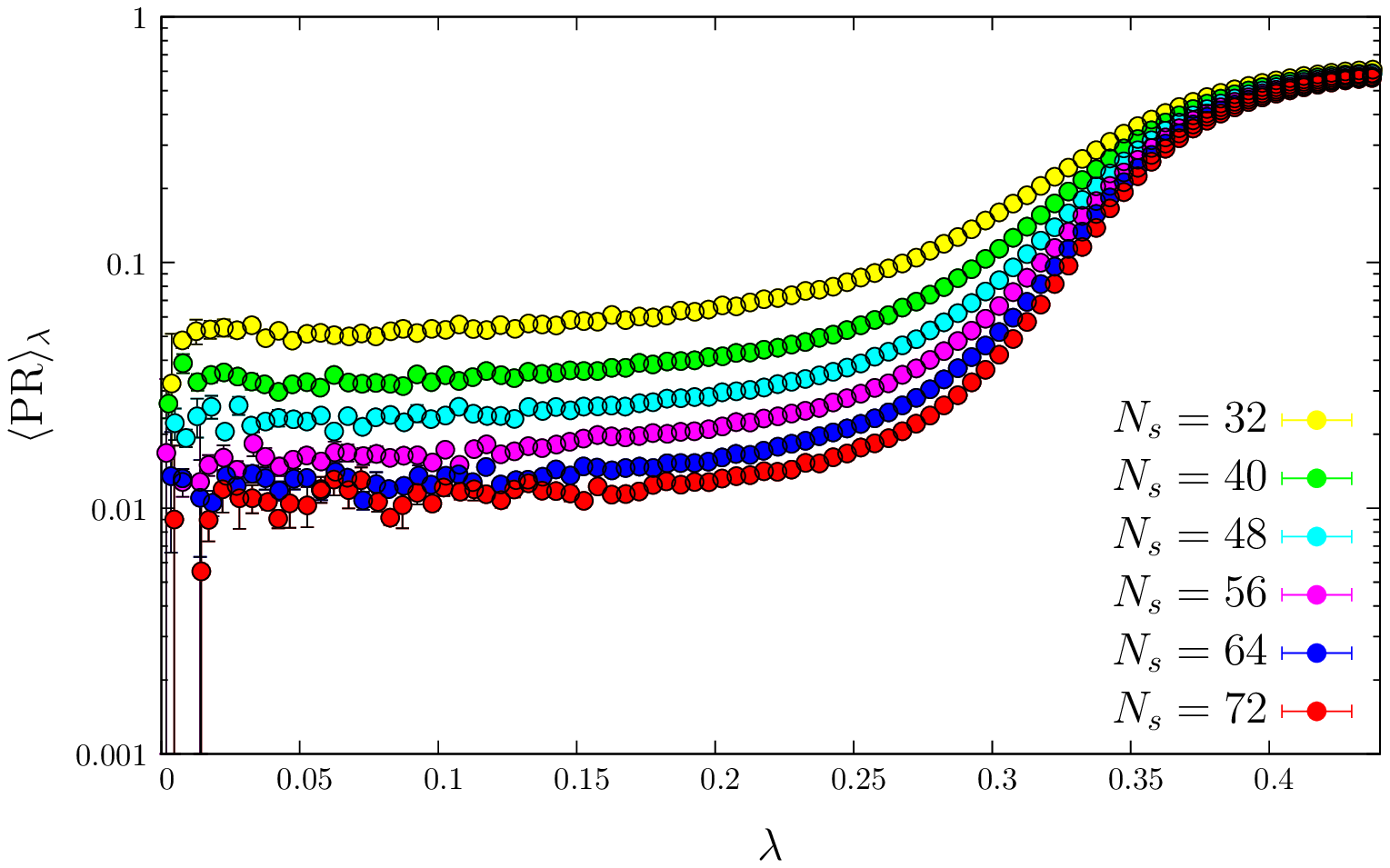}

\vspace{\floatsep}
  \includegraphics[width=0.75\textwidth]{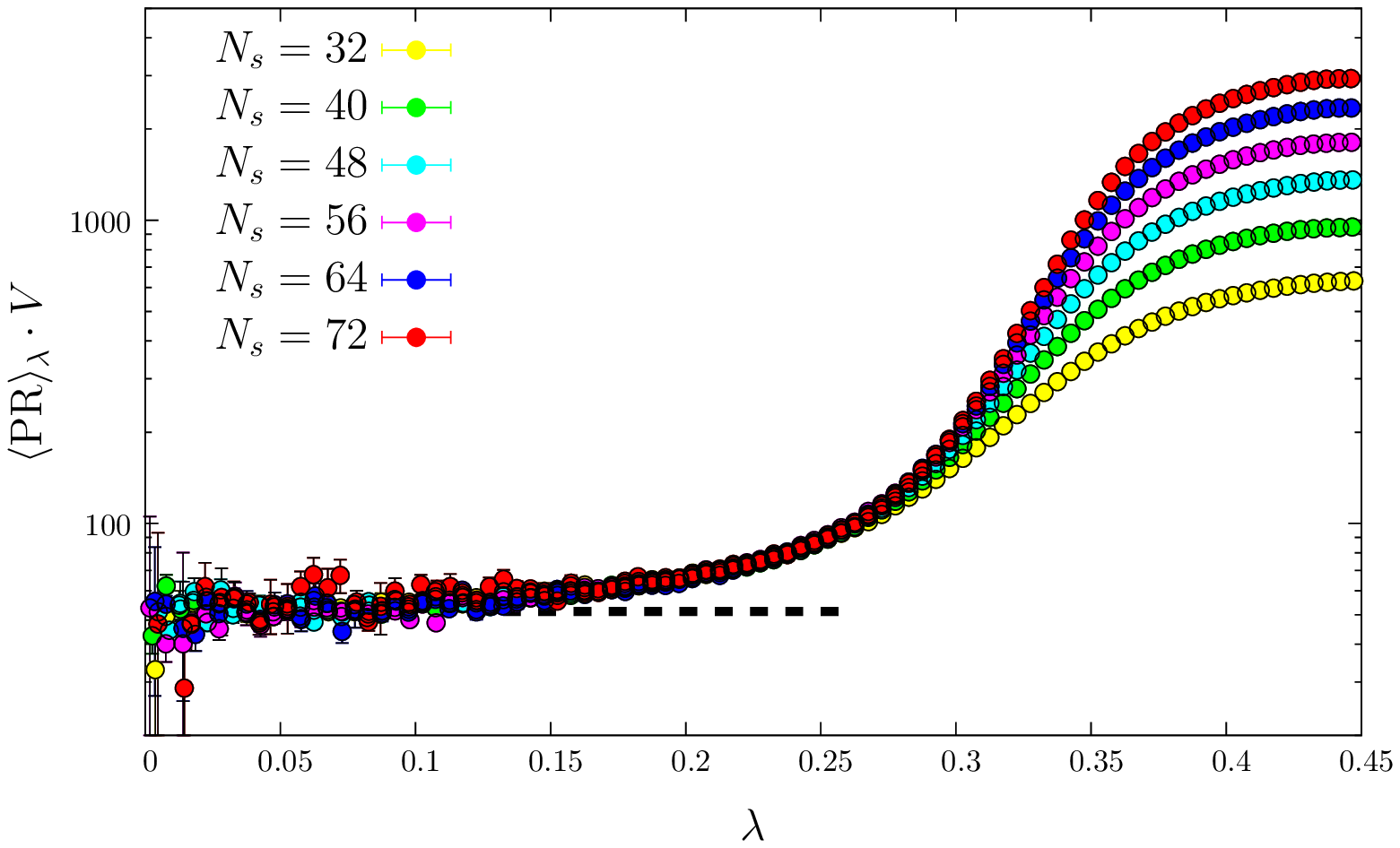}
  \caption{Participation ratio $\la {\rm PR}\ra_\lambda$ (top), and
    size $\la {\rm PR}\ra_\lambda\cdot V$ (bottom) of Dirac
    eigenmodes at $\bred=5.50$ as a function of the spectral region
    for various volumes (in logarithmic scale). The dashed line in the
    bottom panel is the PR of the lowest mode extrapolated to infinite
    lattice size (see text). 
}
  \label{fig:PR}
\end{figure}

\begin{table}[t]
  \centering
  \begin{tabular}{c|c|c}
    $N_s$ & n${}^\circ$ of eigenvalues & n${}^\circ$ of configurations \\
\hline
32 & 100 & 40000 \\
40 & 140 & 40000 \\
48 & 216 & 20000 \\
56 & 280 & 8000 \\
64 & 380 & 4000 \\
72 & 480 & 2000
  \end{tabular}
  \caption{Number of eigenvalues and sample size for the various volumes.}
  \label{tab:details}
\end{table}

I discuss first the PR of the low modes. In Fig.~\ref{fig:PR} I
show how the average PR of the Dirac modes computed locally in the
spectrum, $\la {\rm PR}\ra_\lambda$, changes along the spectrum for
$\bred=5.50$ (in the deconfined phase) and the various lattice
volumes. All locally averaged quantities, like $\la{\rm
  PR}\ra_\lambda$, are obtained by averaging the relevant observable
over the modes in spectral intervals of size $w=0.005$ and over gauge
configurations; the result is assigned to the average eigenvalue
(computed similarly) in that interval. In the top panel of
Fig.~\ref{fig:PR} I show $\la {\rm PR}\ra_\lambda$: it is clear that
this quantity keeps decreasing with the volume for the lowest modes,
while it remains almost constant for the higher modes. In the bottom
panel I show instead $\la {\rm PR}\ra_\lambda \cdot V$, i.e., the
spatial ``size'' of the mode in lattice units: this remains constant
for the lowest modes, and blows up for the higher modes. This shows
that the lowest modes are localised, and the higher ones are
delocalised. Furthermore, the size of the modes is approximately
constant in the localised part of the spectrum. A similar behaviour is
observed at all $\bred$ in the deconfined phase. For the lowest two
values of $\bred$ there, i.e., $\bred=5.05,\,5.10$, the volume
dependence appears to be non-monotonic. This is probably caused by
large finite-size effects near the critical temperature (see below). 

\begin{figure}[t]
  \centering
  \includegraphics[width=0.75\textwidth]{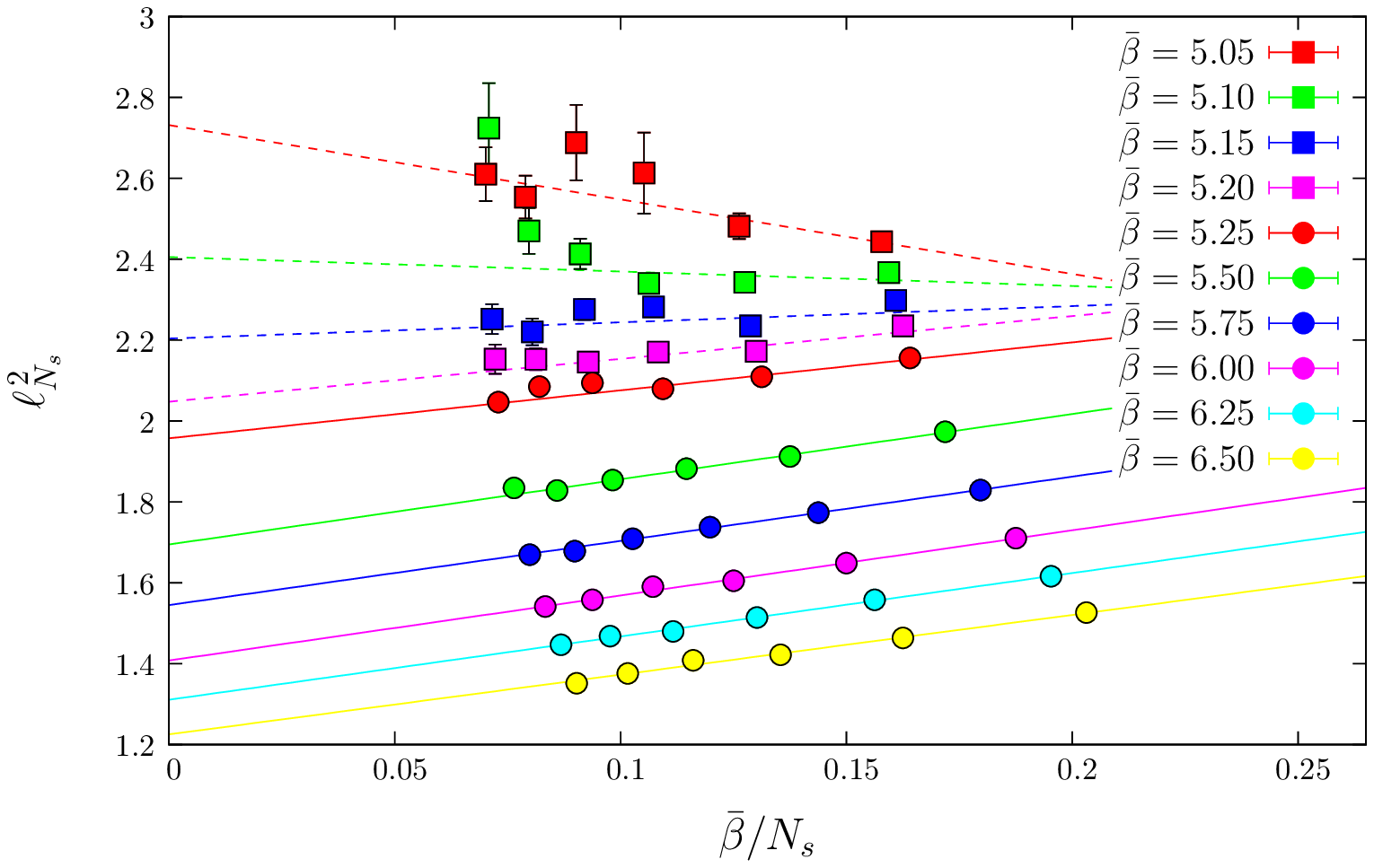}

\vspace{\floatsep}
  \includegraphics[width=0.75\textwidth]{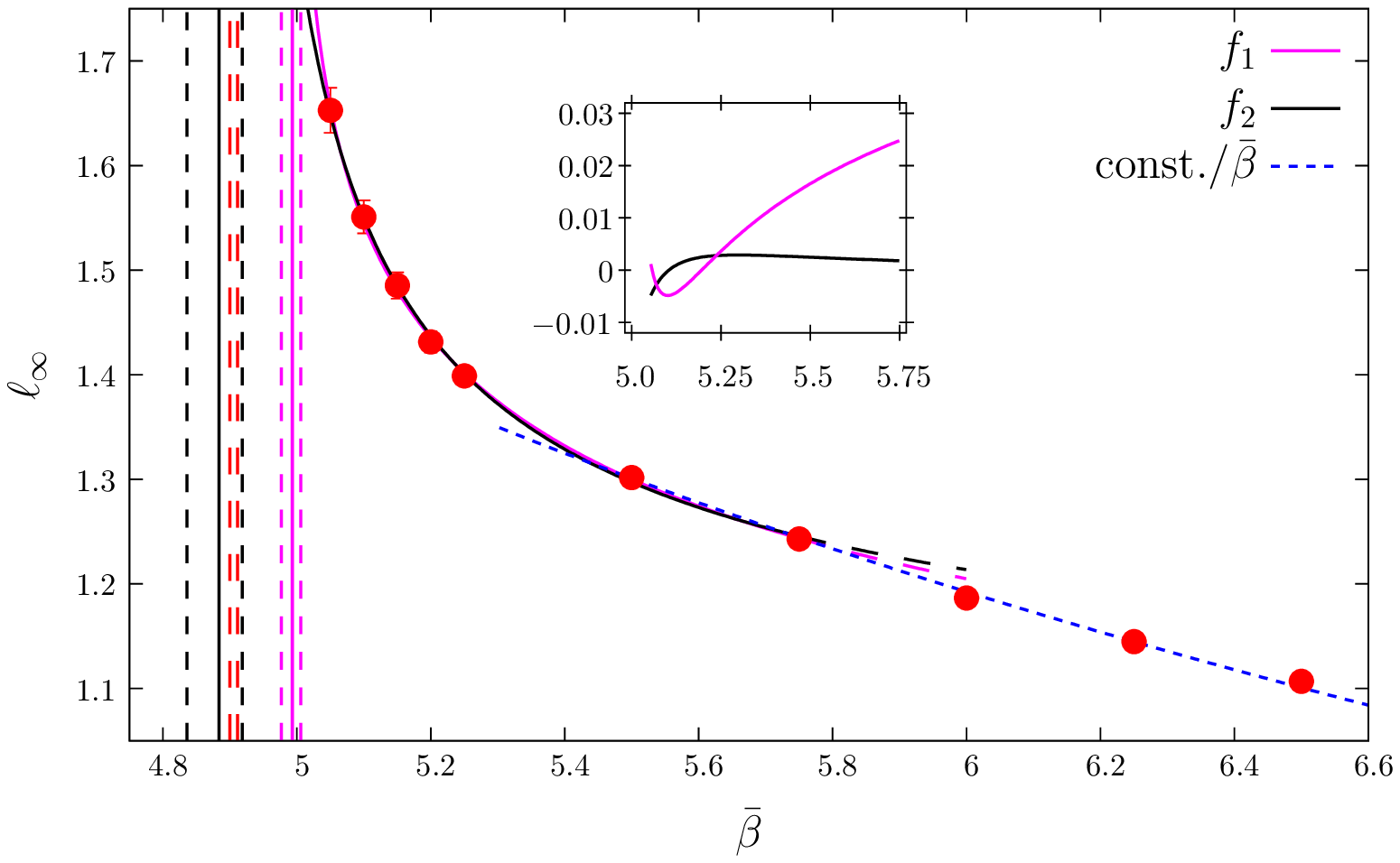}

  \caption{Top panel: size $\ell_{N_s}^2$ of the lowest mode in
    physical units as a function of the inverse linear size of the
    lattice, for various $\bred$ in the deconfined phase. Linear
    fits are also shown with solid lines. Bottom panel: linear size
    $\ell_{\infty}$ extrapolated to infinite volume as a function of
    $\bred$. Fits to the near-$\bred_c$ data with the functions
    $f_{1}(\bred)$ (magenta) and $f_{2}(\bred)$ (black) given in
    Eq.~\protect{\eqref{eq:modesizefit}} are also shown. The
    corresponding fitted values for $\bred_{1,2}$ and the related
    error bands are marked by the vertical lines with corresponding 
    colours. The error band for $\bred_c$ is also shown (red). In the
    inset it is shown the change in the fitting function due to changing the
    fitting interval (see text).
}  
  \label{fig:pr1}
\end{figure}

To give a well-defined estimate of the size of a localised mode in the
deconfined phase, I have measured the PR of the lowest mode averaged over
configurations times the spatial size of the lattice, $\la {\rm
  PR}_1\ra \cdot V$, and extrapolated it to infinite volume. Data
points lie approximately on straight lines when plotted against 
the inverse linear size of the lattice [see Fig.~\ref{fig:pr1} (top)], 
so a linear extrapolation in $1/N_s$ seemed appropriate. The results
compare well with the approximate plateau of $\la {\rm PR}\ra_\lambda
\cdot V$, see the dashed line in Fig.~\ref{fig:PR} (bottom). 
Not surprisingly, this is less so for the lowest two values of
$\bred$. In Fig.~\ref{fig:pr1} (top) I show the size $\ell_{N_s}^2$ of
the lowest mode in physical units\footnote{I use a system of
  ``natural'' units in which $g^2/2=1$, so that the lattice spacing is
  dimensionless.} for the various volumes and values of $\bred$,  
\begin{equation}
  \label{eq:modesize}
  \ell_{N_s}^2 \equiv \la {\rm PR}_1\ra
\left(\f{N_s}{\bred}\right)^2\,.
\end{equation}
The linear extrapolation in $\bred/N_s$ is fully satisfactory for all
values of $\bred$, except for the lowest two, for which the linear fit
to the data is of poorer quality. Nonetheless, the result of the
extrapolation is in agreement with the general trend of the data,
which shows the size of the lowest mode increasing as one gets closer
to the critical coupling. If, as expected, all the modes become
delocalised at the deconfinement point, then this quantity should blow 
up there. Near the critical point, the size of the lowest mode in
lattice units, $\ell_{\infty}\bred$, can then become comparable to
$N_s$ for the available lattice sizes, causing sizeable finite-size
effects. The results for $\ell_{\infty}\bred$ vary from around 7 for
the highest values of $\bred$, to around 8 for the ones closest to
$\bred_c$, so there is probably nothing to worry about for the lattice 
sizes used in this work. On the other hand, if localisation is driven
by the behaviour of the Polyakov line, then near the critical point
one expects further large finite-size effects due to the large
fluctuations of this quantity (see also the discussion below in
Section \ref{sec:numspstat}). In particular, during the numerical
simulations the system can still tunnel between the confined and
deconfined phases for not large enough volumes. Although these effects
are harder to quantify, they are probably responsible for the larger
error bars of $\ell_{N_s}$ near $\bred_c$.   

\begin{table}[t]
  \centering
  \begin{tabular}{c|cc}
    parameter & $i=1$ & $i=2$ \\
\hline
$a_i$     & $1.2058_{-0.0039}^{+0.0040}$  & $0.2045_{-0.0029}^{+0.0037}$  \\
$\bred_i$& $4.993_{-0.016}^{+0.012}$      & $4.884_{-0.048}^{+0.035}$     \\
$c_i$    & $0.1107_{-0.0066}^{+0.0077}$   & $0.498_{-0.048}^{+0.061}$   
  \end{tabular}
  \caption{Results of fits to $\ell_\infty$ with the functional forms
    $f_i(\bred)$ of Eq.~\protect{\eqref{eq:modesizefit}}.
  }
  \label{tab:divloc_ext}
\end{table}

\begin{figure}[t]
  \centering
  \includegraphics[width=0.75\textwidth]{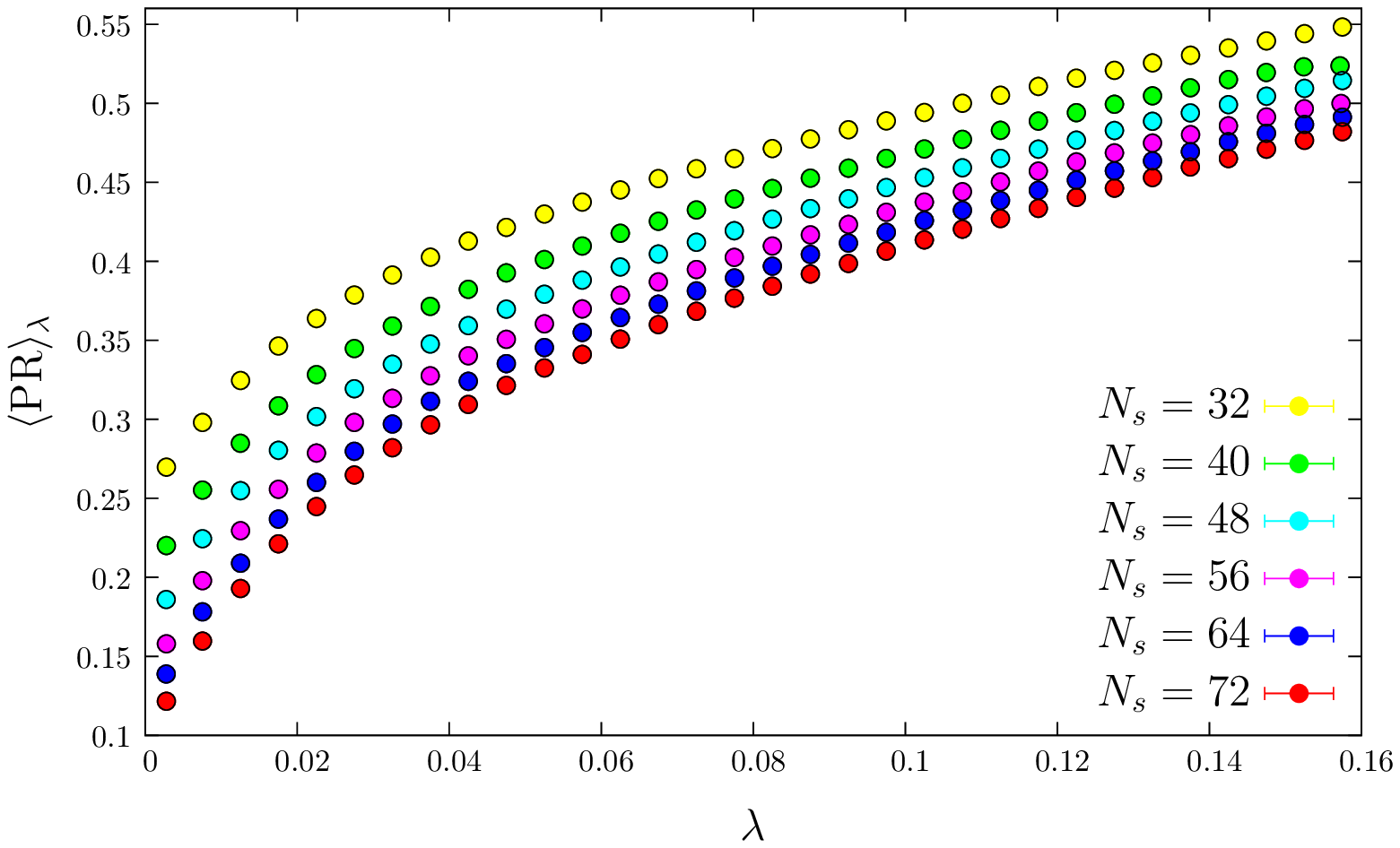}

\vspace{\floatsep}
  \includegraphics[width=0.75\textwidth]{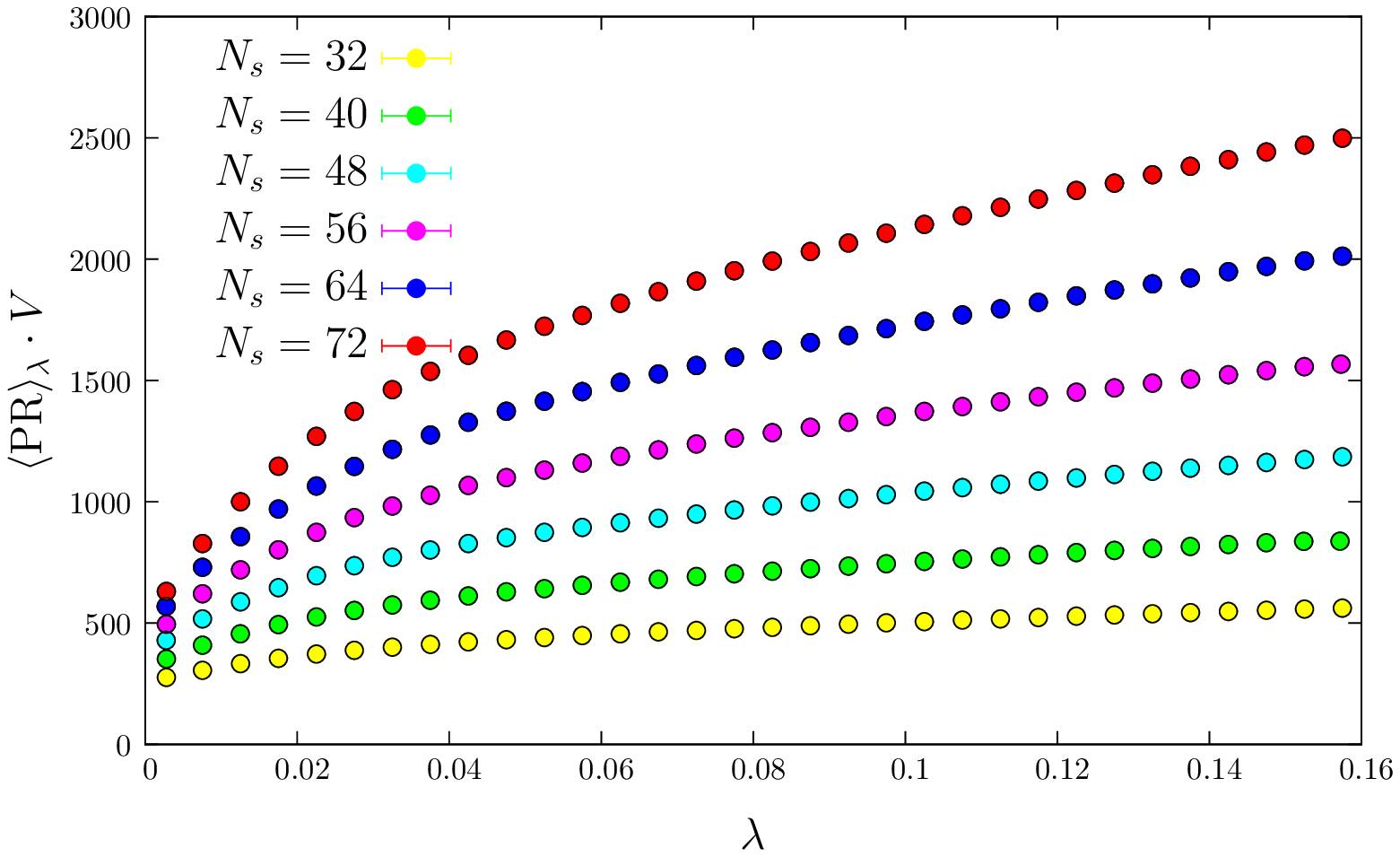}
  \caption{Participation ratio $\la {\rm PR}\ra_\lambda$ (top), 
    and size $\la {\rm PR}\ra_\lambda\cdot V$ (bottom) 
    of Dirac  eigenmodes at $\bred=3.00$ as a function of the spectral
    region for various volumes.}
  \label{fig:PRlow}
\end{figure}

\begin{figure}[t]
  \centering
  \includegraphics[width=0.49\textwidth]{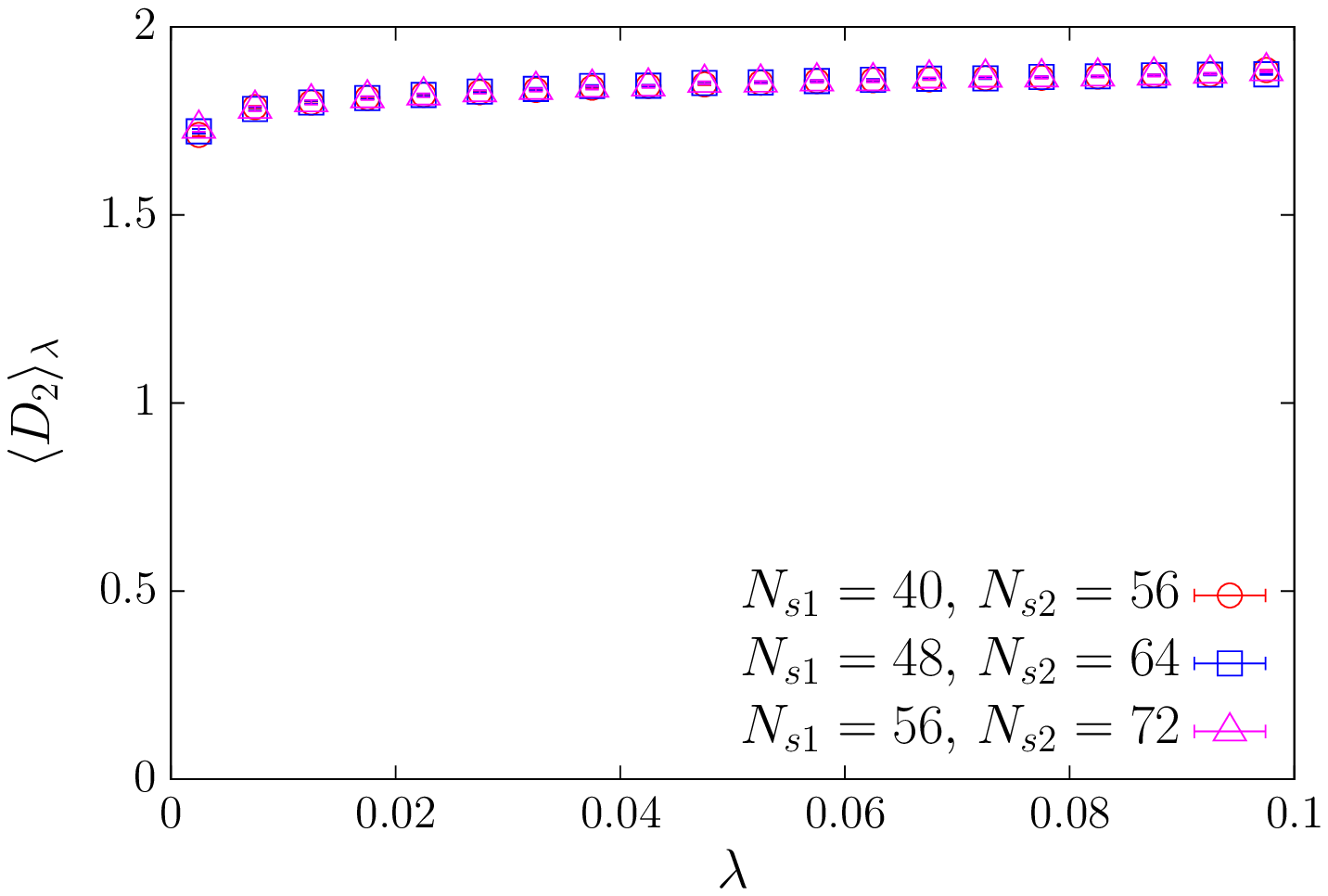}
\hfil
  \includegraphics[width=0.49\textwidth]{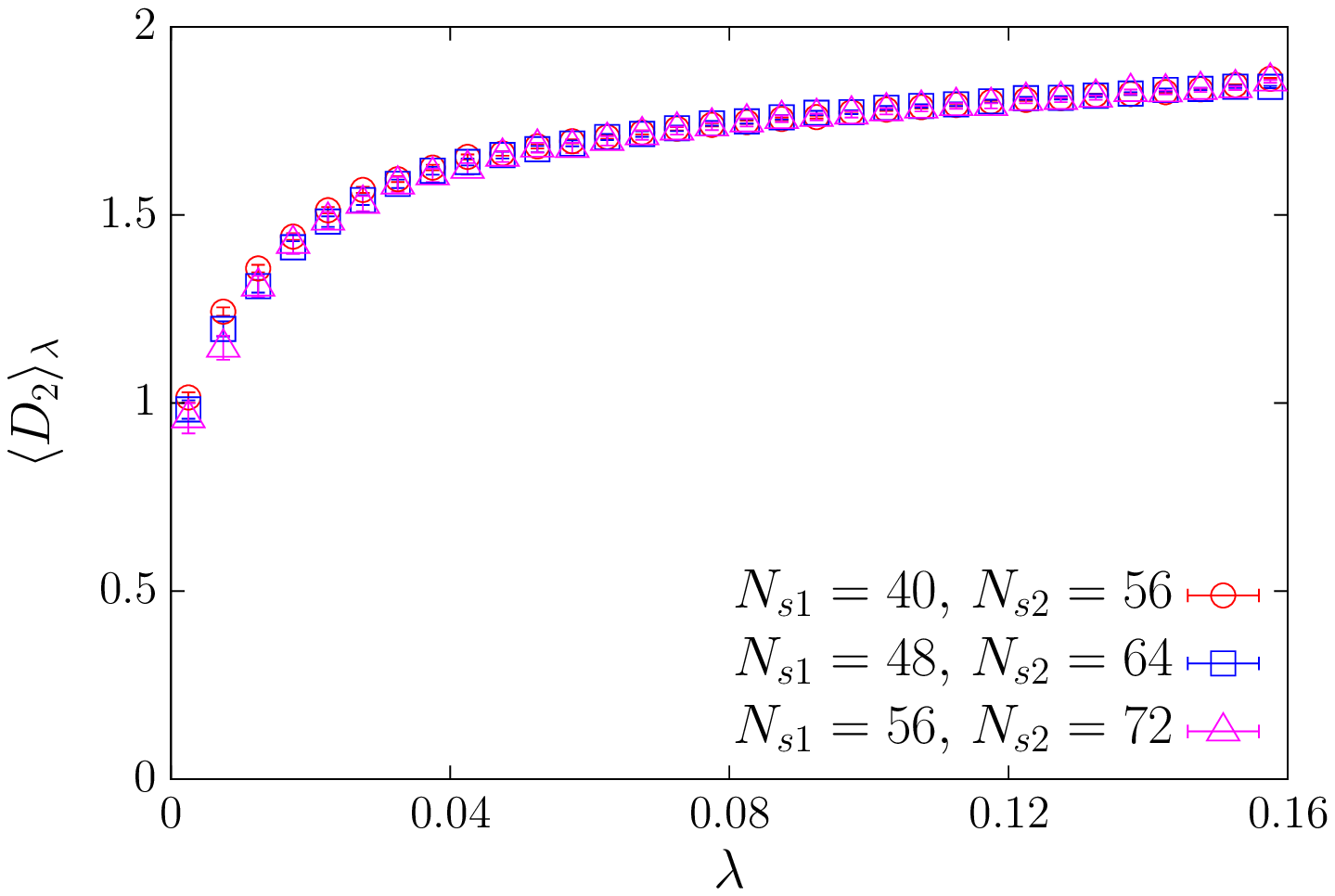}

\vspace{\floatsep}
  \includegraphics[width=0.49\textwidth]{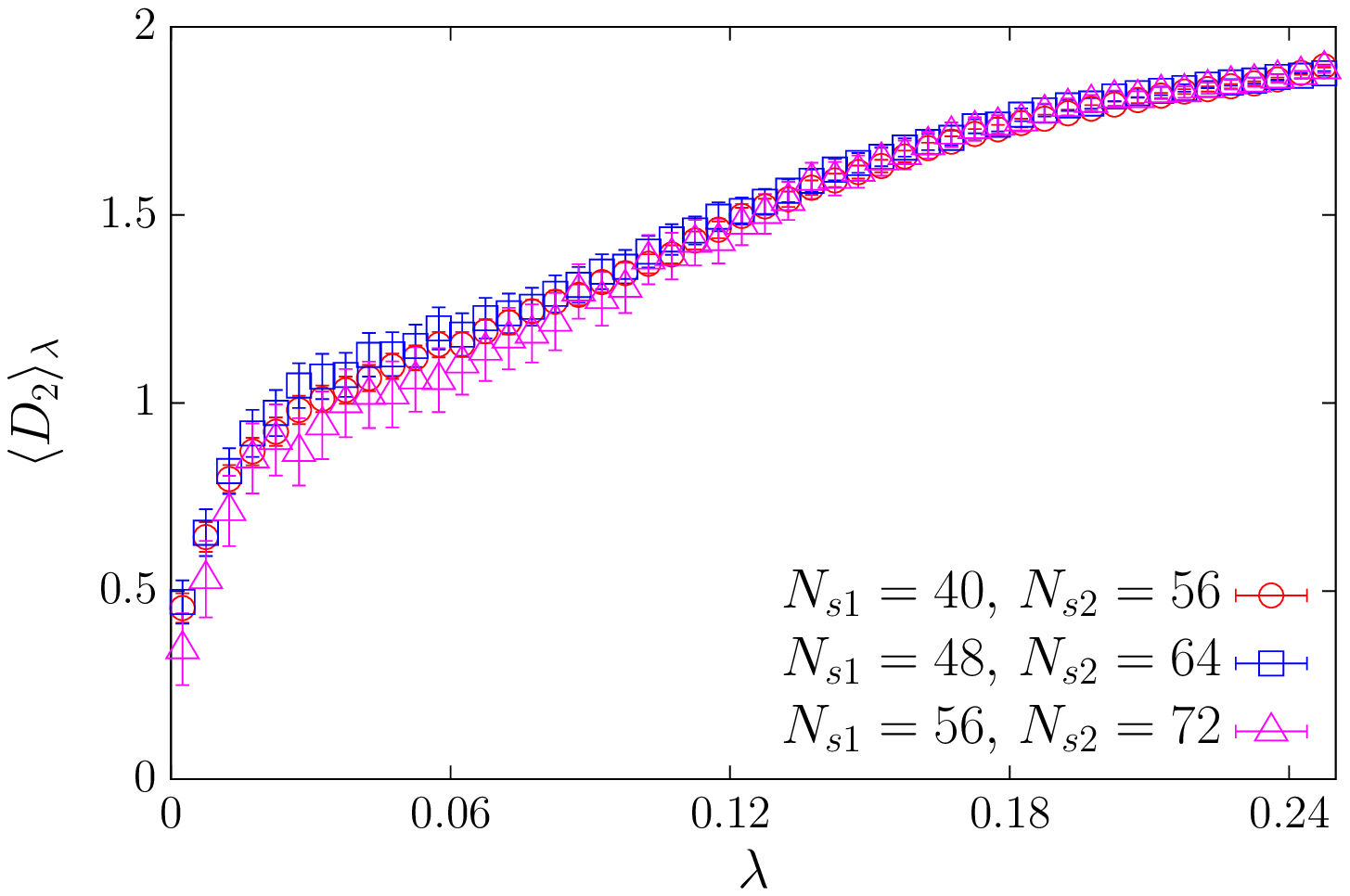}
  \caption{Fractal dimension of the Dirac eigenmodes as a function of
    the spectral region in the confined phase, for $\bred=2.50$
    (top left), $\bred=3.00$ (top right), and $\bred=4.00$ (bottom). 
  }
  \label{fig:dim2}
\end{figure}

The results of the infinite-volume extrapolation of $\ell_{N_s}$ are
shown in Fig.~\ref{fig:pr1} (bottom). For the lowest values of
$\bred$, even though one may not fully trust quantitatively the
results due to the above-mentioned finite-size effects, there is a
clear sign of a divergence as the critical coupling is approached. At
large $\bred$ instead this quantity decreases as $1/\bred$, reflecting
the fact that there the extrapolated size of the lowest mode in
lattice units is almost independent of the coupling. A constant fit
using only $\bred\ge 5.50$ yields $\ell_{\infty}\bred = 7.155(10)$
with $\chi^2/{\rm d.o.f.}=1.4$. This leads to a linear dependence of
the physical linear size of the mode on the inverse temperature at
high temperature, as one would expect given that this is the most
important scale there. The same behaviour has been observed in 
QCD~\cite{Kovacs:2012zq}. Plugging in the fit result, one gets
$\ell_{\infty}T= 1.7887(25)$, where $T\equiv\bred/N_t$.
Despite the limited number of points close to $\bred_c$, I have tried to
fit the divergent behaviour with the following functional forms,
\begin{equation}
  \label{eq:modesizefit}
  f_1(\bred)=\f{a_1}{(\bred-\bred_1)^{c_1}}\,, \qquad
  f_2(\bred)=\exp\left\{\f{a_2}{(\bred-\bred_2)^{c_2}}\right\}\,.
\end{equation}
The first one is a simple power-law, while the second one is
inspired by the behaviour of the localisation length\footnote{It must
  be remarked that in general the localisation length, $\xi$,
  and the quantities $\ell^{(q)} = ({\rm IPR}_q)^{-\f{1}{2(q-1)}}$
  built out of the generalised IPRs (see footnote \ref{foot:genIPR}),
  do not have the same critical behaviour. In fact, even though all
  these quantities provide an estimate of the size of the localised
  modes and should blow up at the critical point, they are sensitive
  to different features of the modes, which usually results in
  different critical exponents.} at criticality.\footnote{The critical 
  behaviour of $\xi$ should not depend on whether the
  critical line is crossed in the direction of the eigenvalues at
  fixed $\bred$, or in the direction of $\bred$ at fixed
  eigenvalue. Having extrapolated the lowest mode to infinite volume,
  here we are looking at around $\lambda=0$.}  
Only points up to $\bred=5.75$ were included. Both fits converge, with
suspiciously small  $\chi^2/{\rm d.o.f.}\simeq 0.2$ and $\chi^2/{\rm
  d.o.f.}\simeq 0.5$, respectively, indicating a fair amount of
overfitting. The values obtained for the delocalisation point,
$\bred_{1,2}$, are reported in Table \ref{tab:divloc_ext}. While
$\bred_1$ is inconsistent with the critical lattice coupling
$\bred_c=4.9057(57)$~\cite{Liddle:2008kk}, $\bred_2$ is in fair
agreement with it. To check which functional form describes better the 
data, I restricted the fit to points up to $\bred=5.25$ only. Again
both fits converge, with an even bigger amount of overfitting
($\chi^2/{\rm d.o.f.}\simeq 0.1$ and $\chi^2/{\rm d.o.f.}\simeq 0.08$,
respectively), and increased statistical errors on the parameters. The
important point, however, is that while $f_1$ changes visibly in the
region left out of the fit (up to 2\%), $f_2$ changes very little
(always less than 0.2\%). I take this as an indication that $f_2$
captures better the singular behaviour of $\ell_\infty$. Despite all
the limitations of the present analysis, one can quite safely conclude
that the results are compatible with localised modes appearing at
$\bred_c$ in the thermodynamic limit. Quite interestingly, the result
for the critical exponent $c_2$ found using $f_2(\bred)$ is
surprisingly close to the critical exponent of the localisation length 
$\nu=\f{1}{2}$. Even though no firm conclusion can be reached yet
based on this result, it suggests nonetheless the possibility that the
generalised IPRs (see footnote \ref{foot:genIPR}) may vanish
exponentially rather than as power laws as the critical point is
approached.  

The behaviour of the PR in the confined phase is different, although
apparently still nontrivial. In Fig.~\ref{fig:PRlow} I show 
$\la {\rm PR}\ra_\lambda$ and $\la {\rm PR}\ra_\lambda \cdot V$ for
$\bred=3.00$ for the various lattice sizes. None of these two
quantities shows a clear sign of converging to a constant anywhere in
the available spectrum, indicating that they show a somewhat
intermediate behaviour between localised and fully delocalised. To
investigate this issue quantitatively, one can compute the fractal
dimension $D_2$, related to the PR as~\cite{Evers:2008zz}
\begin{equation}
  \label{eq:dim2}
  {\rm PR} \sim L^{D_2-d} = L^{D_2-2}\,,
\end{equation}
where $\sim$ denotes the asymptotic large-volume behaviour and $L$ is
the linear size of the system. For fully delocalised modes $D_2=2$,
while for localised modes $D_2=0$. The fractal dimension $D_2$ can be
estimated by comparing the PR computed on pairs of lattices of linear
spatial sizes $N_{s1}$ and $N_{s2}$, using the formula
\begin{equation}
  \label{eq:dim2_2}
  D_2(N_{s1},N_{s2}) = 2 + \f{\log({\rm PR}(N_{s1})/{\rm
      PR}(N_{s2}))}{\log(N_{s1}/N_{s2})}\,. 
\end{equation}
From this estimator one obtains $D_2$ in the limit of large
$N_{s1,2}$. The results for $\bred=2.50$, $3.00$, $4.00$ are shown in  
Fig.~\ref{fig:dim2} for three different pairs of volumes. The volume
dependence is mild to non-existent. Low modes have a nontrivial
fractal dimension between 0 and 2, while for higher modes $D_2$
approaches 2. Moreover, low modes are closer and closer to being
localised as one approaches the critical temperature from below. 

\begin{figure}[t]
  \centering
  \includegraphics[width=0.75\textwidth]{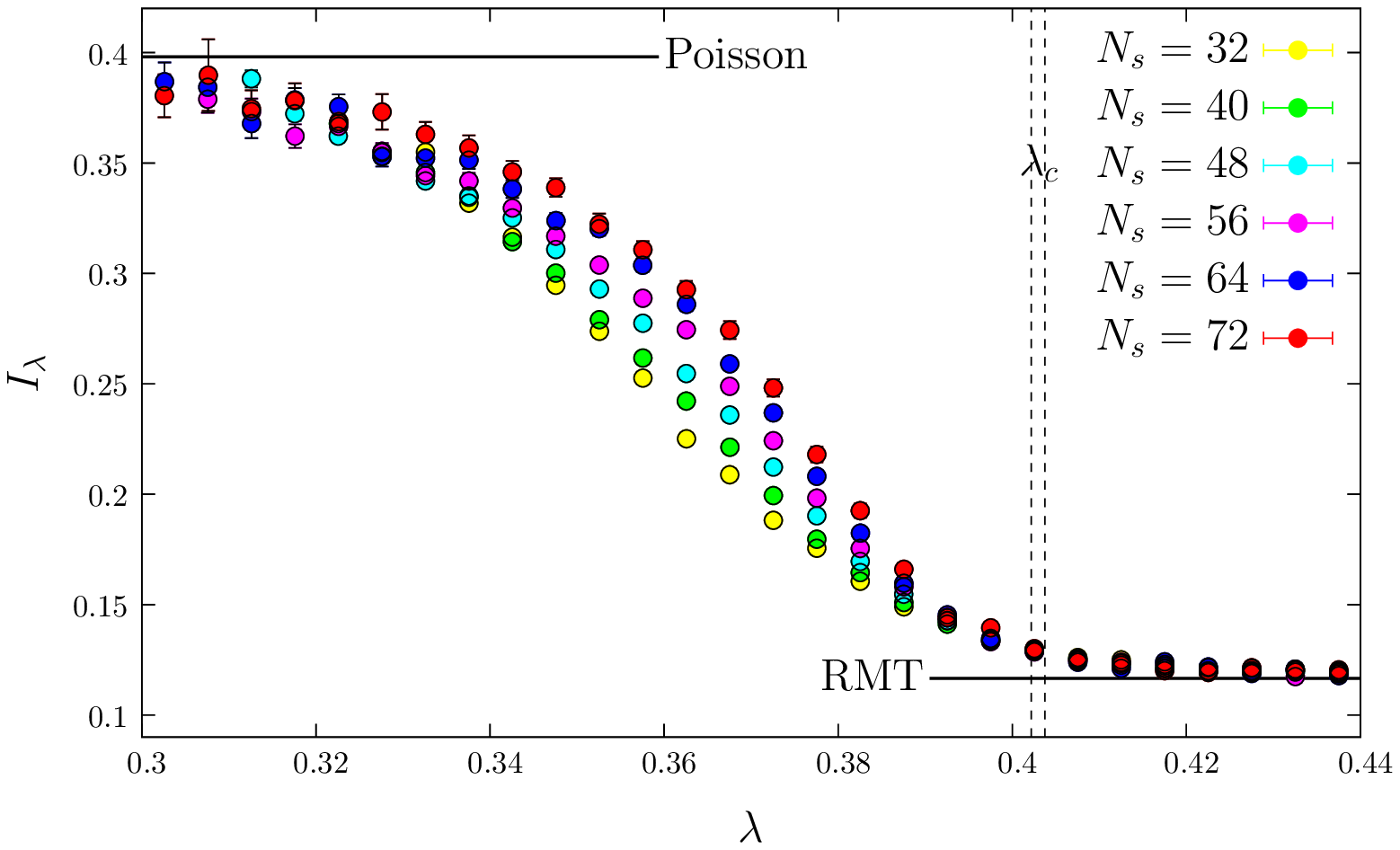}

\vspace{\floatsep}
  \includegraphics[width=0.75\textwidth]{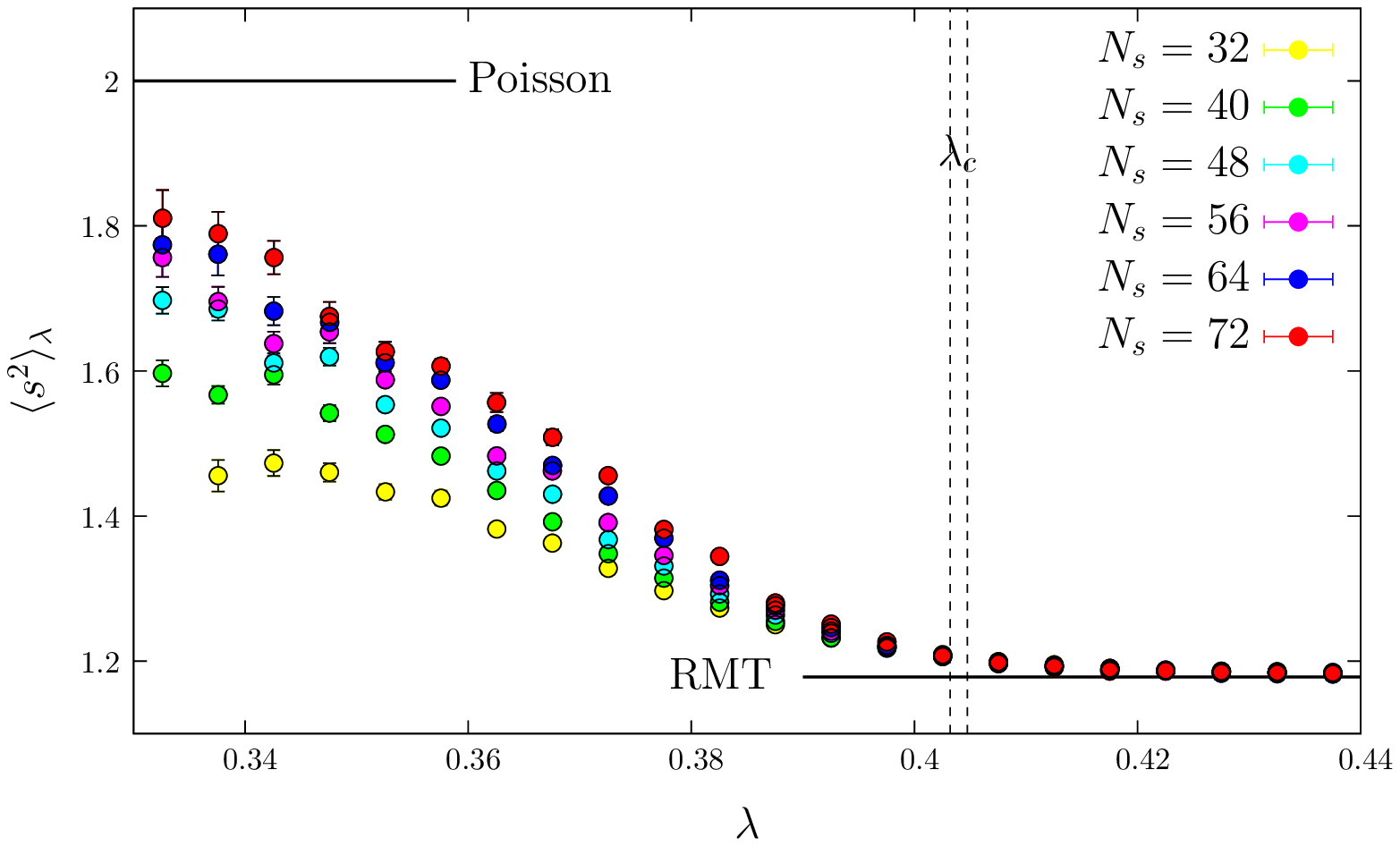}
  \caption{Dependence on the spectral region for $I_\lambda$ (top) and
    $\la s^2\ra_\lambda$ (bottom) in the deconfined phase at
    $\bred=6.25$ for several volumes. Only points with $\la
    s\ra_\lambda$ close to 1 are shown. The position of the mobility 
    edge and its uncertainty, as determined via a finite-size scaling
    analysis, are also shown.}  
  \label{fig:spstat}
\end{figure}

\subsection{Spectral statistics}
\label{sec:numspstat}

\begin{figure}[t]
  \centering
  \includegraphics[width=0.49\textwidth]{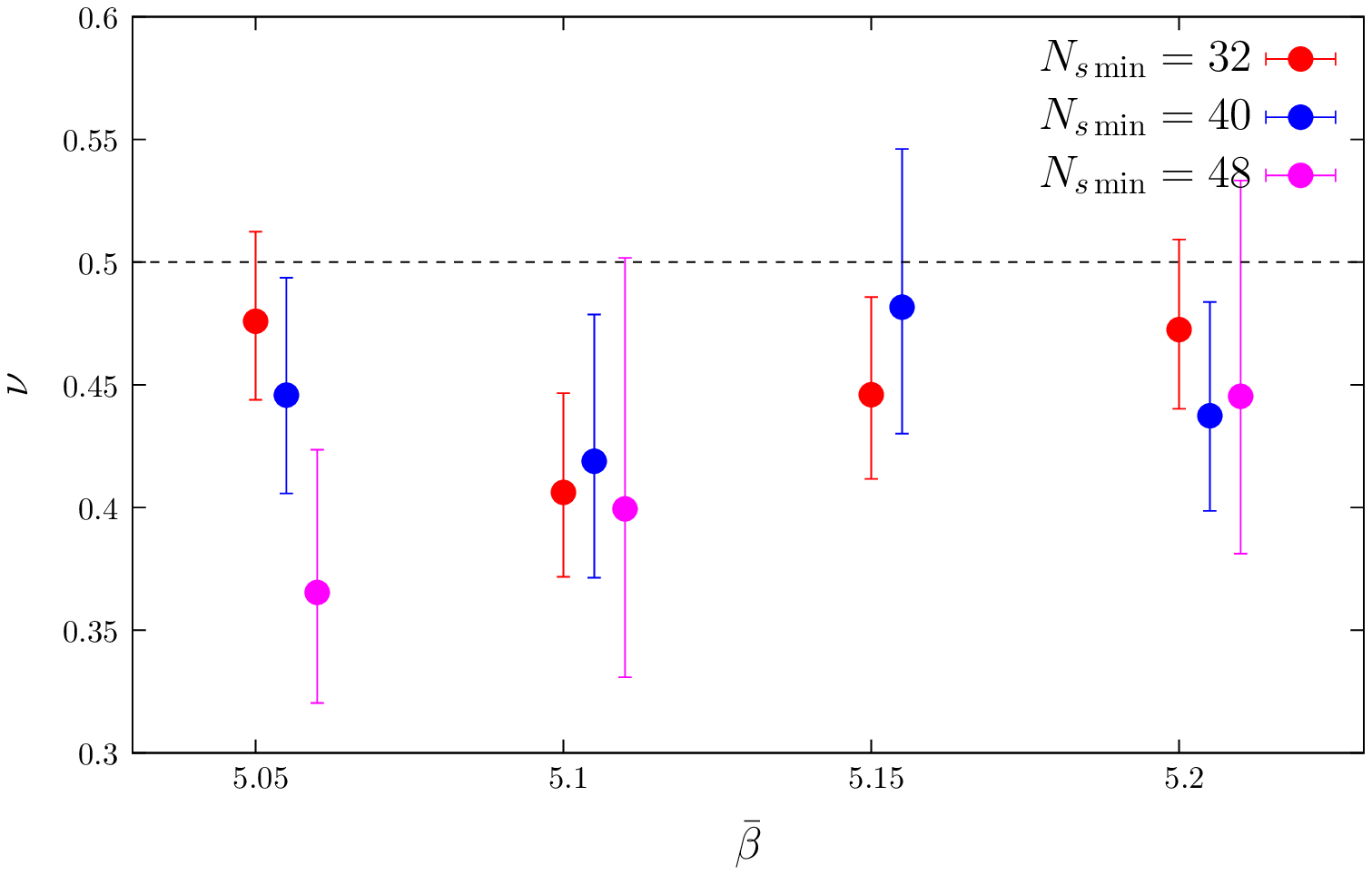}\hfil  \includegraphics[width=0.49\textwidth]{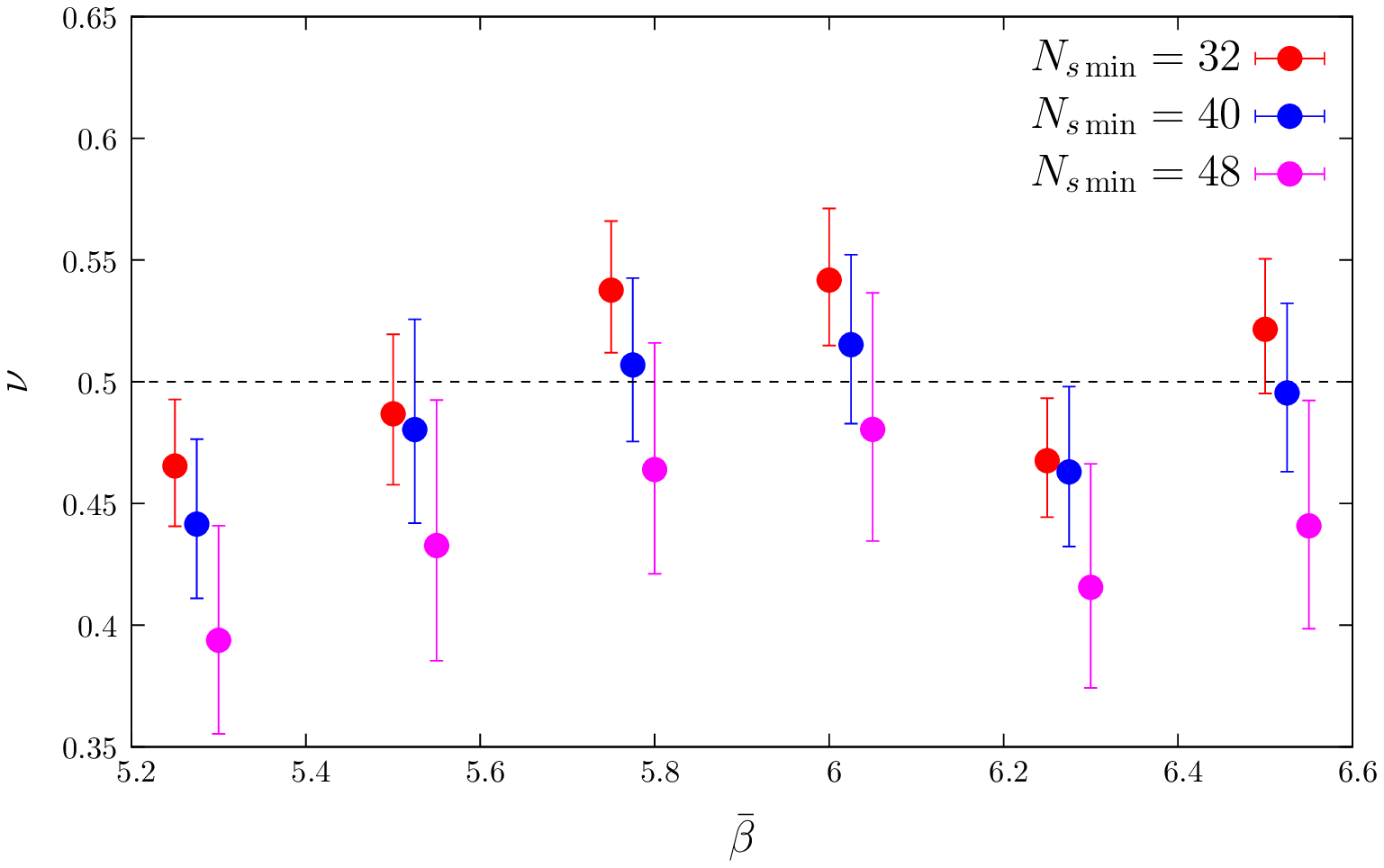}
  \caption{Critical exponent $\nu$ obtained via an unconstrained fit
    with the function Eq.~\eqref{eq:sc_func_unc} to the data for the
    spectral statistic $I_\lambda$ in the deconfined phase, close 
    to the transition (left panel) and farther above the transition
    (right panel).}
  \label{fig:nuunc}
\end{figure}

As mentioned above in Section \ref{sec:loclgt}, spectral statistics
can be used to study the localisation properties of the Dirac modes
and to determine efficiently the  mobility edge and the related
critical properties. The typical behaviour of $I_\lambda$ and $\la
s^2\ra_\lambda$ in the deconfined phase is shown in
Fig.~\ref{fig:spstat} for $\bred=6.25$. Low modes are close to having
Poisson statistics, and more and more so as the size of the lattice
increases. For every lattice size the statistics change from Poisson
to RMT-type, more precisely to GUE-type, as one moves up along the
spectrum. At some point the curves corresponding to the various
volumes merge, and show little to no dependence on the volume. This is
precisely the behaviour expected for a disorder-driven BKT transition
in the spectrum. To make this statement quantitative, I verified the
scaling hypothesis Eq.~\eqref{eq:BKT_scaling} by fitting the data for
$I_\lambda$ for various volumes with the rational function
\begin{equation}
  \label{eq:sc_func_unc}
  \begin{aligned}
    F(\lambda,N_s) &= \f{c_1 + c_2 y(\lambda,N_s) + c_3
      y(\lambda,N_s)^2 + c_4 y(\lambda,N_s)^3}{1 + \bar{c}_1
      y(\lambda,N_s) +     \bar{c}_2 y(\lambda,N_s)^2}\,, \\
    y(\lambda,N_s)&=(\lambda-\lambda_c)(\log N_s)^{\f{1}{\nu}}\,.
  \end{aligned}
\end{equation}
The fit was performed restricting to an interval of width $w=0.06$
centred at the merging point of the curves, and including volumes
with $N_s\ge N_{s\,{\rm min}}$ for $N_{s\,{\rm min}}=32,40,48$. I
obtained reasonable values of $\chi^2/{\rm d.o.f.}$, ranging between 
$1$ and $2.15$ for $N_{s\,{\rm  min}}=32$, between $0.5$ and $1.05$
for $N_{s\,{\rm min}}=40$, and between $0.35$ and $0.65$ for
$N_{s\,{\rm min}}=48$. The results for the critical exponent $\nu$,
shown in Fig.~\ref{fig:nuunc}, are in fair agreement with the
theoretically expected value $\nu=\f{1}{2}$. 

\begin{figure}[t]
  \centering
  \includegraphics[width=0.49\textwidth]{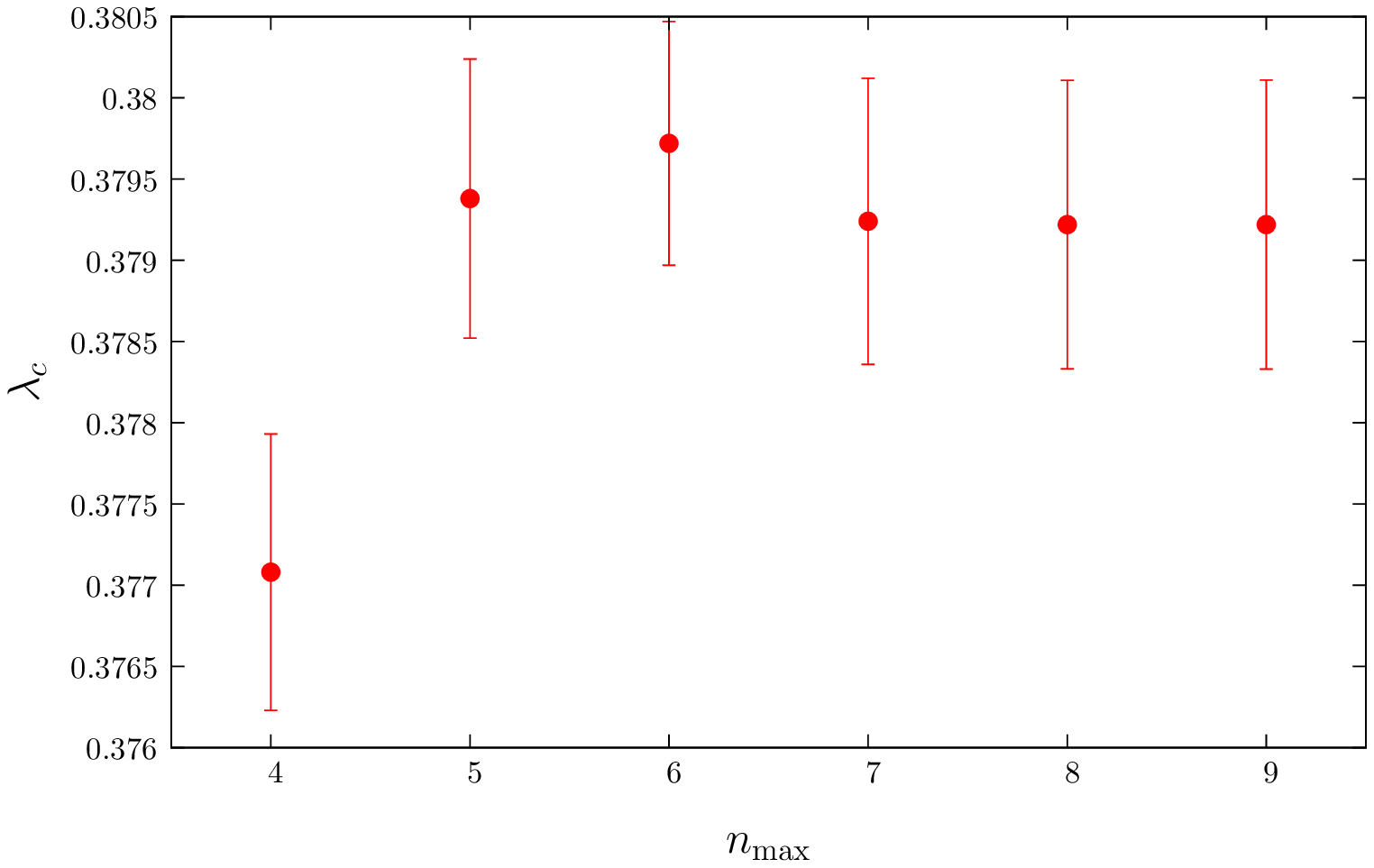}\hfil  \includegraphics[width=0.49\textwidth]{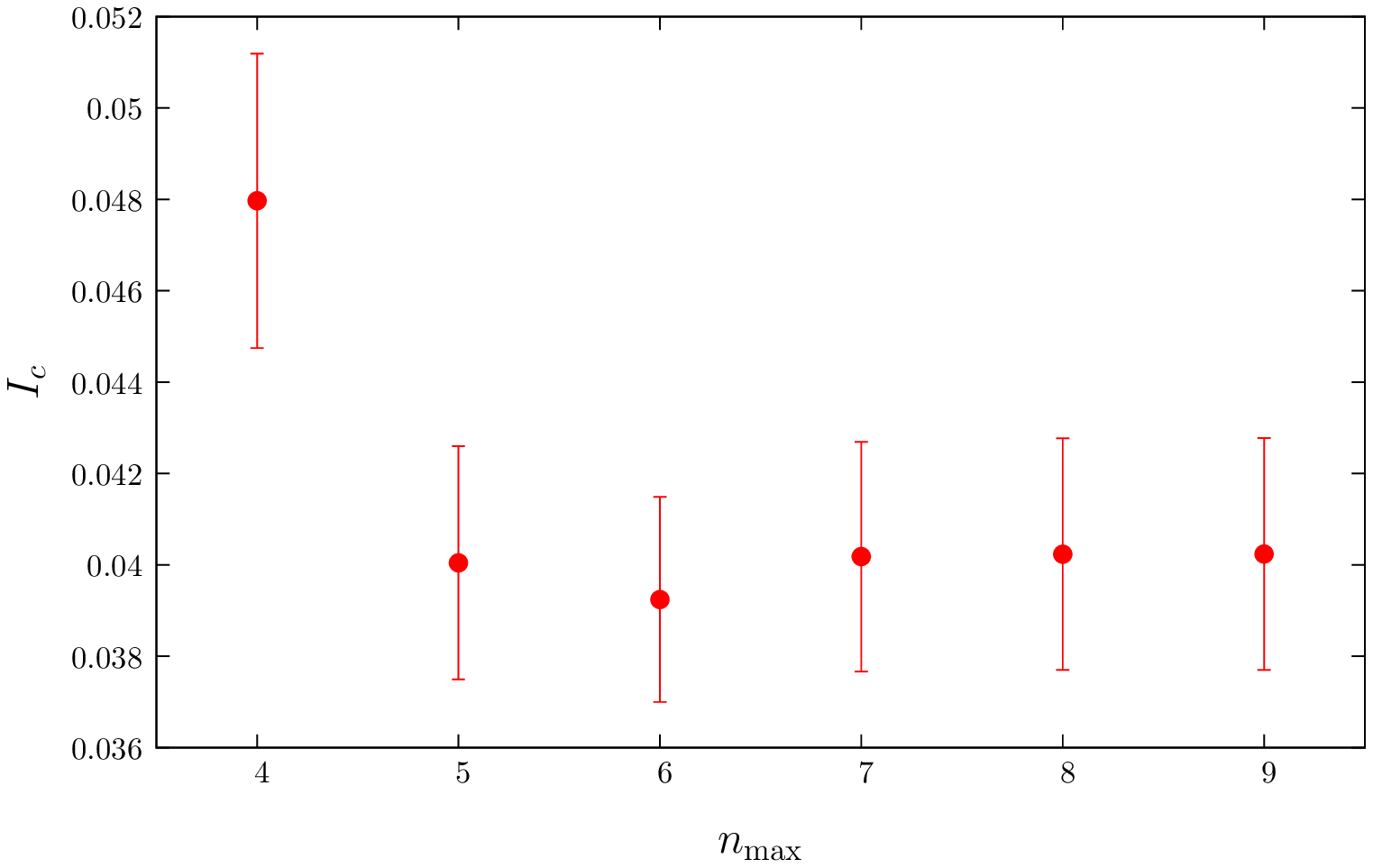} 
  \caption{Dependence of the fitted values of $\lambda_c$ (left) and
    $I_c$ (right) on the order of the polynomial in a constrained fit
    to the data for the spectral statistic $I_\lambda$ obtained at
    $\bred=5.75$, using $w=0.06$, $N_{s\,{\rm min}}=40$ and $\sigma=8$.}
  \label{fig:constr}
\end{figure}

\begin{figure}[t]
  \centering
\includegraphics[width=0.75\textwidth]{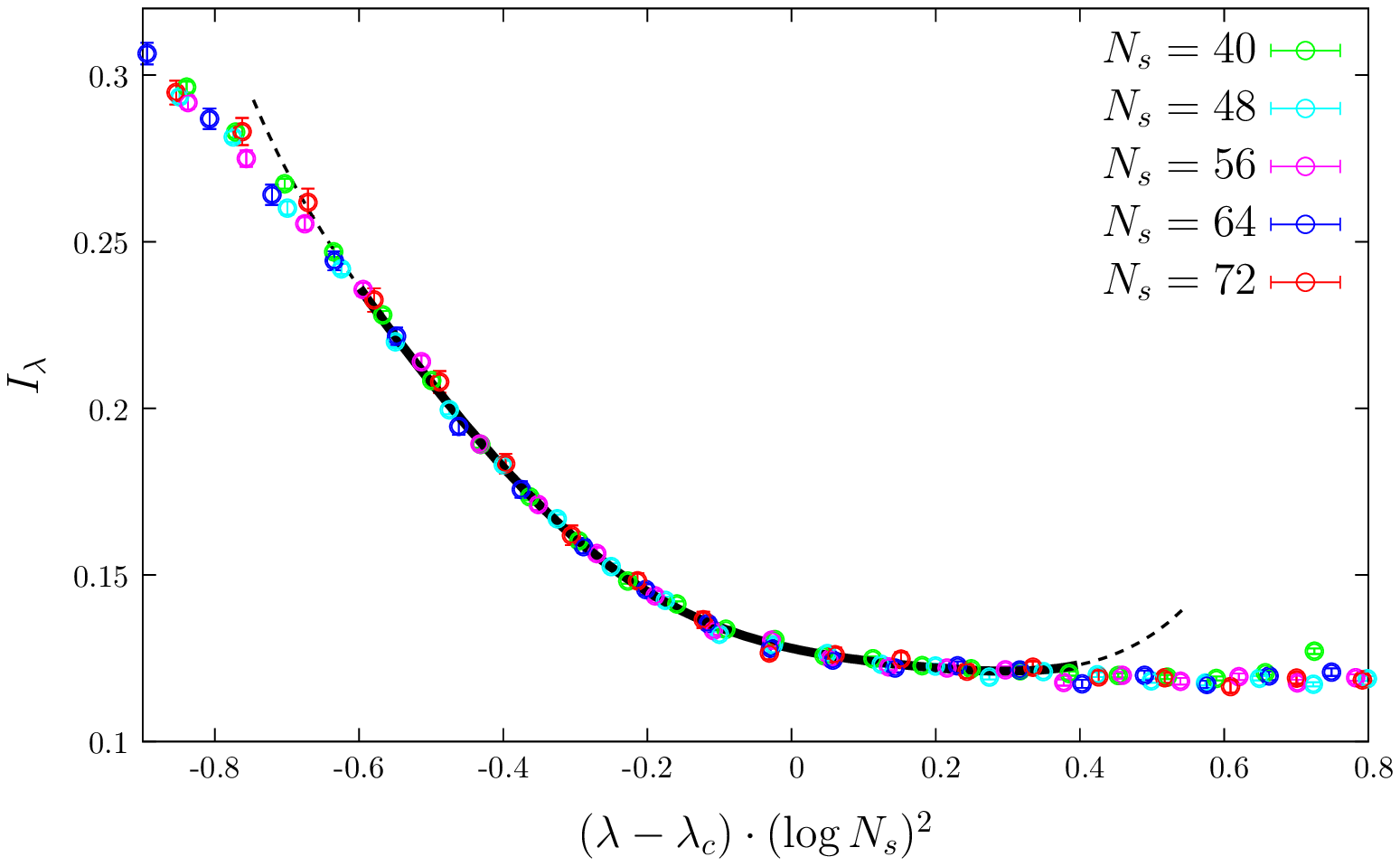}

\vspace{\floatsep}
\includegraphics[width=0.75\textwidth]{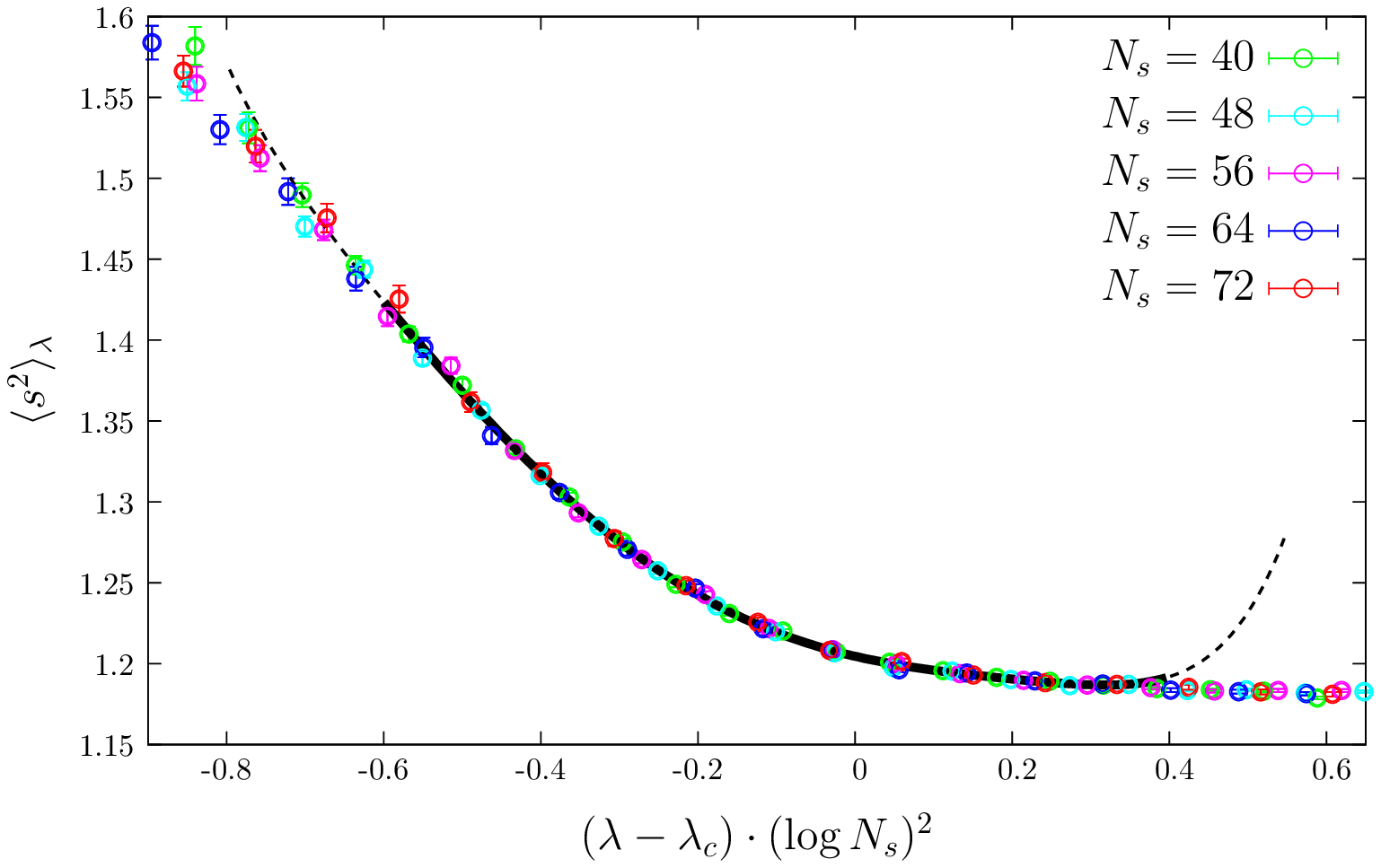}
  \caption{Scaling behaviour of $I_\lambda$ (top) and $\la
    s^2\ra_\lambda$ (bottom) at $\bred=5.75$. The scaling function
    (solid line) was obtained via constrained fitting using $w=0.06$,
    $N_{s\,{\rm min}}=40$ and $\sigma=8$. The scaling behaviour is
    seen to persist also outside the range where the fit was
    performed, where the fitted function does not match the data
    points (dashed line).} 
  \label{fig:scal}
\end{figure}

Fits to the data with Eq.~\eqref{eq:sc_func_unc} are quite sensitive
to the initial values of the fitting parameters and the choice of
fitting interval, and sometimes the error estimate of the parameters
provided by MINOS is not accurate, especially close to the critical
coupling (the missing point at $\bred=5.15$ for $N_{s\,{\rm min}}=48$ in
Fig.~\ref{fig:nuunc} is precisely a case of inaccurate error
estimate). Despite these numerical shortcomings, the results are all
quite consistent with the hypothesis $\nu=\f{1}{2}$. In fact, I
performed these fits to obtain an unbiased (or at least as little
biased as possible) estimate of the critical exponent, as a
preliminary step for a more accurate and more systematic finite-size 
scaling study using constrained fits~\cite{Lepage:2001ym}. These fits
turn out to be unstable if $\nu$ is not constrained strongly, for
which reason I preferred to first check the viability of the
theoretical value with unconstrained fits, and then perform the
constrained fits fixing $\nu=\f{1}{2}$. Constrained fits were done by
fitting the data with polynomial functions, 
\begin{equation}
  \label{eq:sc_func_constr}
  F(\lambda,N_s) = \sum_{n=0}^{n_{\rm max}} F_n\, y(\lambda,N_s)^n\,,
\end{equation}
where $y(\lambda,N_s)$ is given in Eq.~\eqref{eq:sc_func_unc}, which
is nothing but a truncated Taylor expansion around $\lambda_c$. The
coefficients $F_n$ and the mobility edge $\lambda_c$ are the fitting
parameters. In general, in a constrained fit one minimises an
augmented $\chi^2$, $\chi^2_{\rm aug} = \chi^2 + \chi^2_{\rm prior}$,
where 
\begin{equation}
  \label{eq:prior}
  \chi^2_{\rm prior} =
  \f{\left(\lambda_c-\lambda_c^{(0)}\right)^2}{\sigma_{\lambda_c}^2} +
  \f{\left(\nu-\nu^{(0)}\right)^2}{\sigma_\nu^2 } + \sum_{n=0}^{n_{\rm
      max}}\f{\left(F_n-F_n^{(0)}\right)^2}{\sigma_n^2} \,,
\end{equation}
with properly chosen priors $\lambda_c^{(0)},\,\nu^{(0)},\,F_n^{(0)}$
and $\sigma_{\lambda_c},\,\sigma_{\nu},\,\sigma_n$, reflecting one's 
knowledge of the parameters. The order $n_{\rm max}$ of the polynomial
is then increased until the error on the parameters stabilises. This
allows to estimate accurately the systematic error due to
truncation~\cite{Lepage:2001ym}. In my analysis I used no prior for the
first four coefficients and for the mobility edge (so formally setting
$\sigma_{\lambda_c}=\sigma_1=\sigma_2=\sigma_3=\sigma_4=\infty$), I
fixed $\nu=\f{1}{2}$ (so formally setting $\sigma_\nu=0$), and set
$F_n^{(0)}=0$ and a rather loose and constant width $\sigma_n=\sigma$
for $n\ge 5$. I repeated the analysis for $\sigma=3,5,8$, with little
variation of the results. I used spectral intervals of width
$w=0.055$ and $w=0.06$ around the merging point of the curves, and
included data for $N_s\ge N_{s\,{\rm min}}$ with $N_{s\,{\rm
    min}}=40,48$. This was done to check further sources of 
systematic error. The order $n_{\rm max}$ of the polynomial was
increased up to $n_{\rm max} = 9$. The typical results of the
constrained fitting procedure are shown in Fig.~\ref{fig:constr},
where I show $\lambda_c$ and the value $I_c\equiv I_{\lambda_c}= F_0$
of $I_\lambda$ at the mobility edge, obtained by fitting $I_\lambda$ for
$\bred=5.75$, $w=0.06$, $N_{s\,{\rm min}}=40$ and $\sigma=8$. The
convergence of the central values and of the errors is clear. 
The procedure was then repeated for the spectral statistic $\la
s^2\ra_\lambda$. The quality of the scaling can be seen in
Fig.~\ref{fig:scal}. 

\begin{figure}[t]
  \centering
  \includegraphics[width=0.49\textwidth]{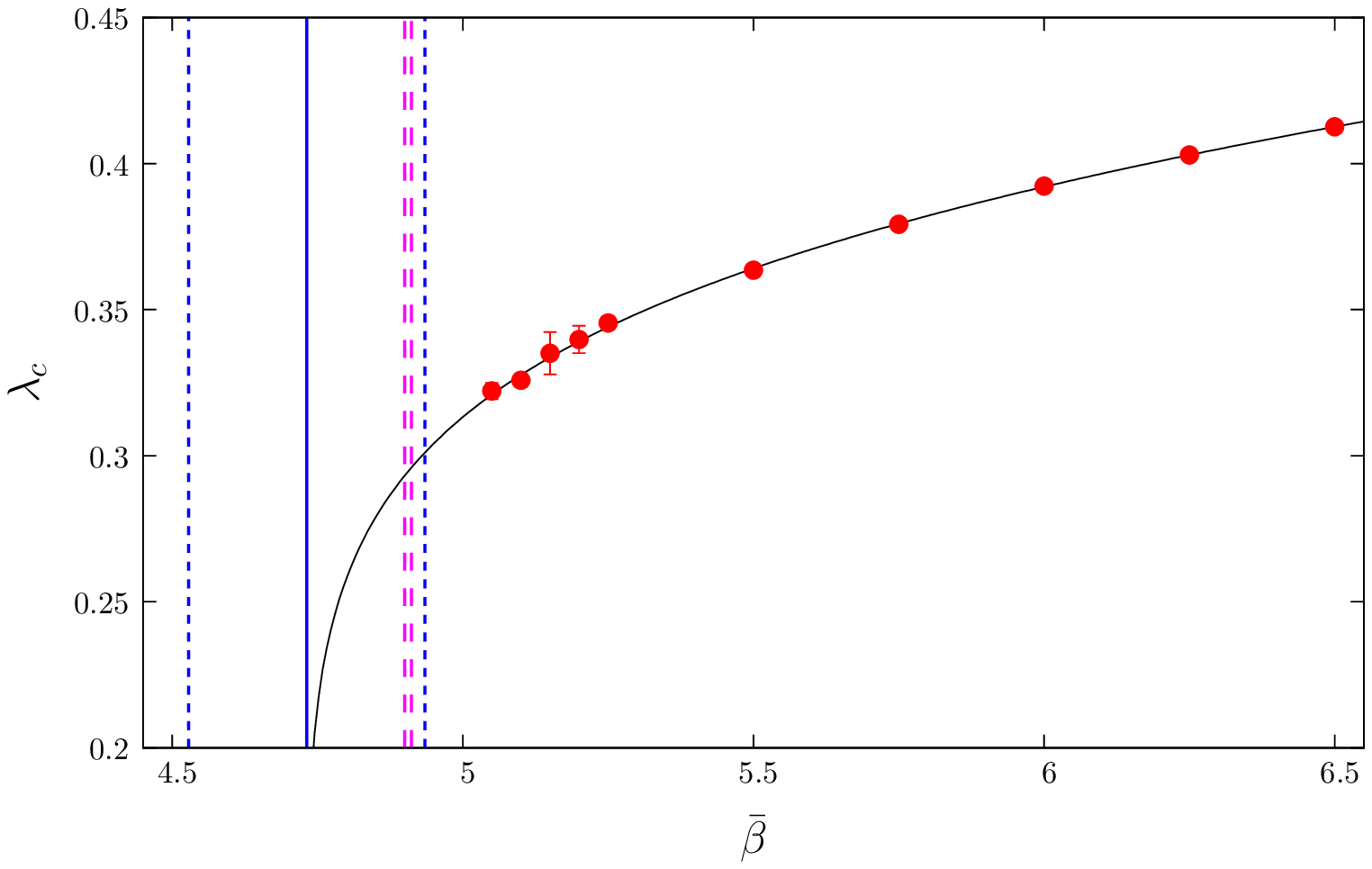}\hfil  \includegraphics[width=0.49\textwidth]{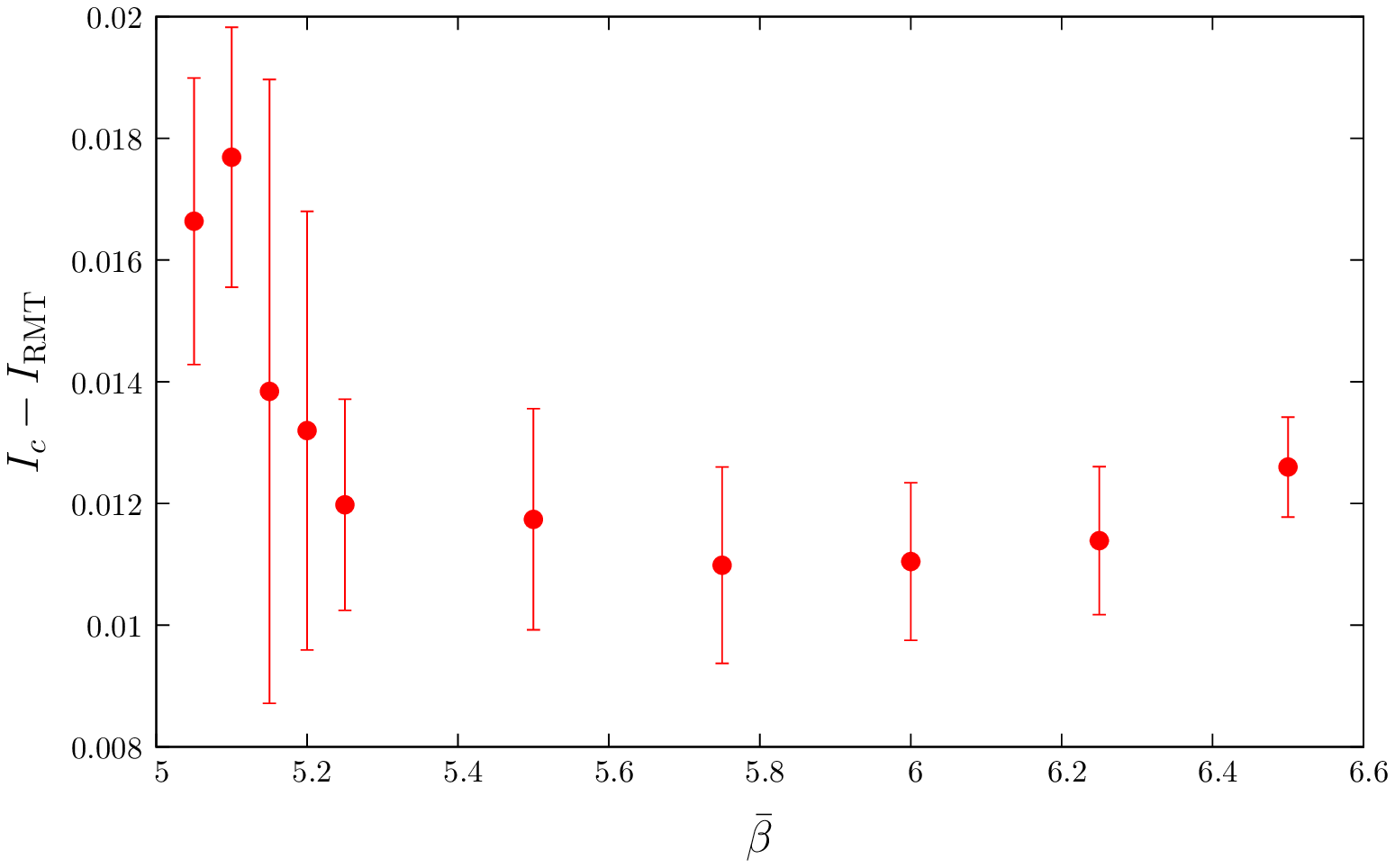} 
  \caption{Mobility edge $\lambda_c$ (in lattice units) obtained via
    constrained fitting using $I_\lambda$ (left), and deviation of the
    corresponding value $I_c$ of the statistic at the critical point
    from the RMT value (right). The result of a fit of $\lambda_c$
    with Eq.~\eqref{eq:lambdac_beta} is also shown (solid line),
    together with the value $\bred_0$ at which the  mobility edge
    extrapolates to zero and corresponding error band (blue lines),
    and with the error band for the critical lattice coupling for
    deconfinement obtained in Ref.~\protect{\cite{Liddle:2008kk}}
    (magenta lines).}     
  \label{fig:crit_I}
\end{figure}

\begin{figure}[t]
  \centering
  \includegraphics[width=0.49\textwidth]{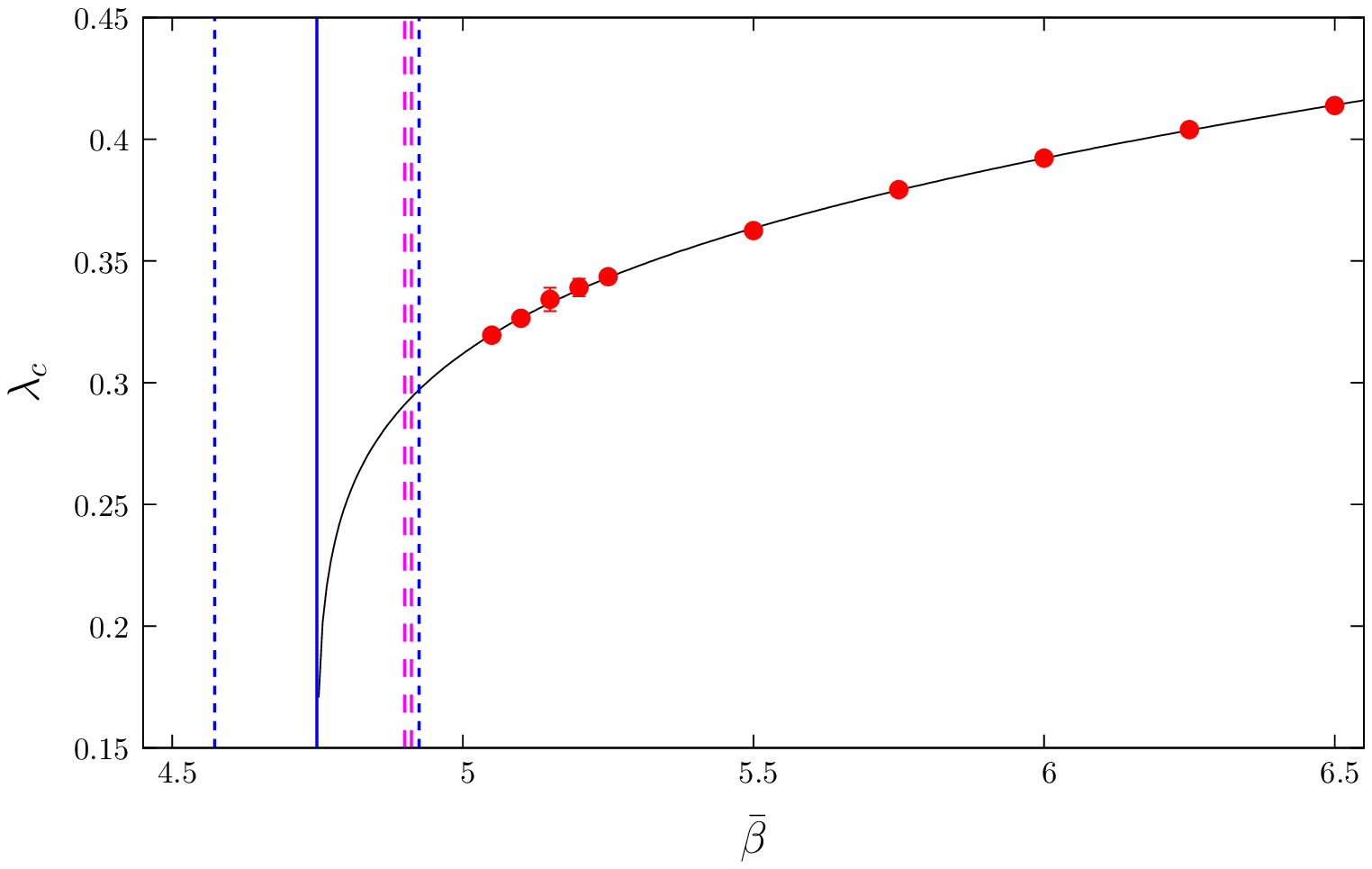}\hfil  \includegraphics[width=0.49\textwidth]{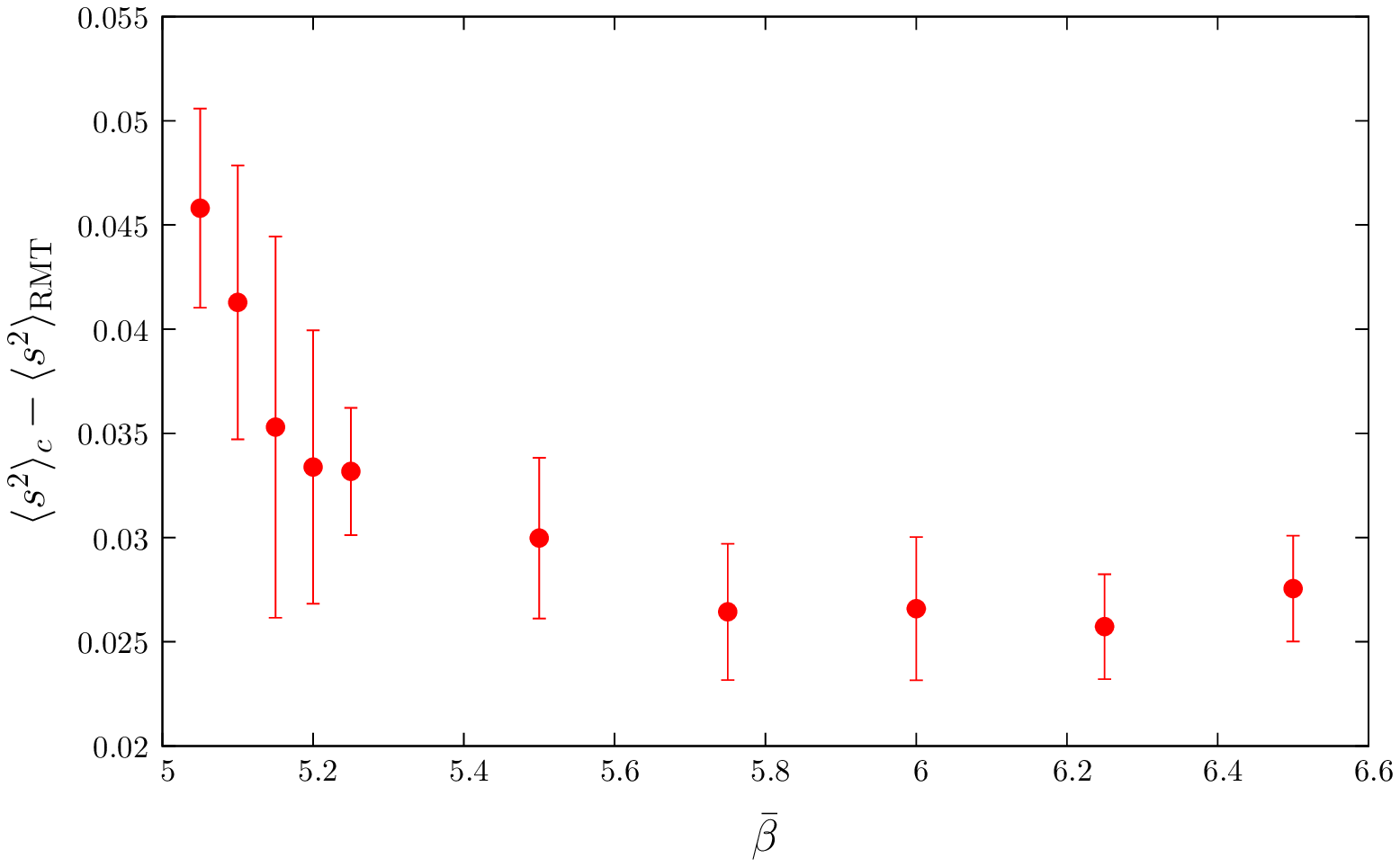} 
  \caption{Mobility edge $\lambda_c$ (in lattice units) obtained via
    constrained fitting using $\la s^2\ra_\lambda$ (left), and
    deviation of the corresponding value $\la s^2\ra_c$ of the
    statistic at the critical point from the RMT value (right). The
    result of a fit of $\lambda_c$ with Eq.~\eqref{eq:lambdac_beta} is
    also shown (solid line), together with the value $\bred_0$ at
    which the  mobility edge extrapolates to zero and corresponding
    error band (blue lines), and with the error band for the critical
    lattice coupling for deconfinement obtained in
    Ref.~\protect{\cite{Liddle:2008kk}} (magenta lines).}    
  \label{fig:crit_s2}
\end{figure}

\begin{figure}[t]
  \centering
  \includegraphics[width=0.75\textwidth]{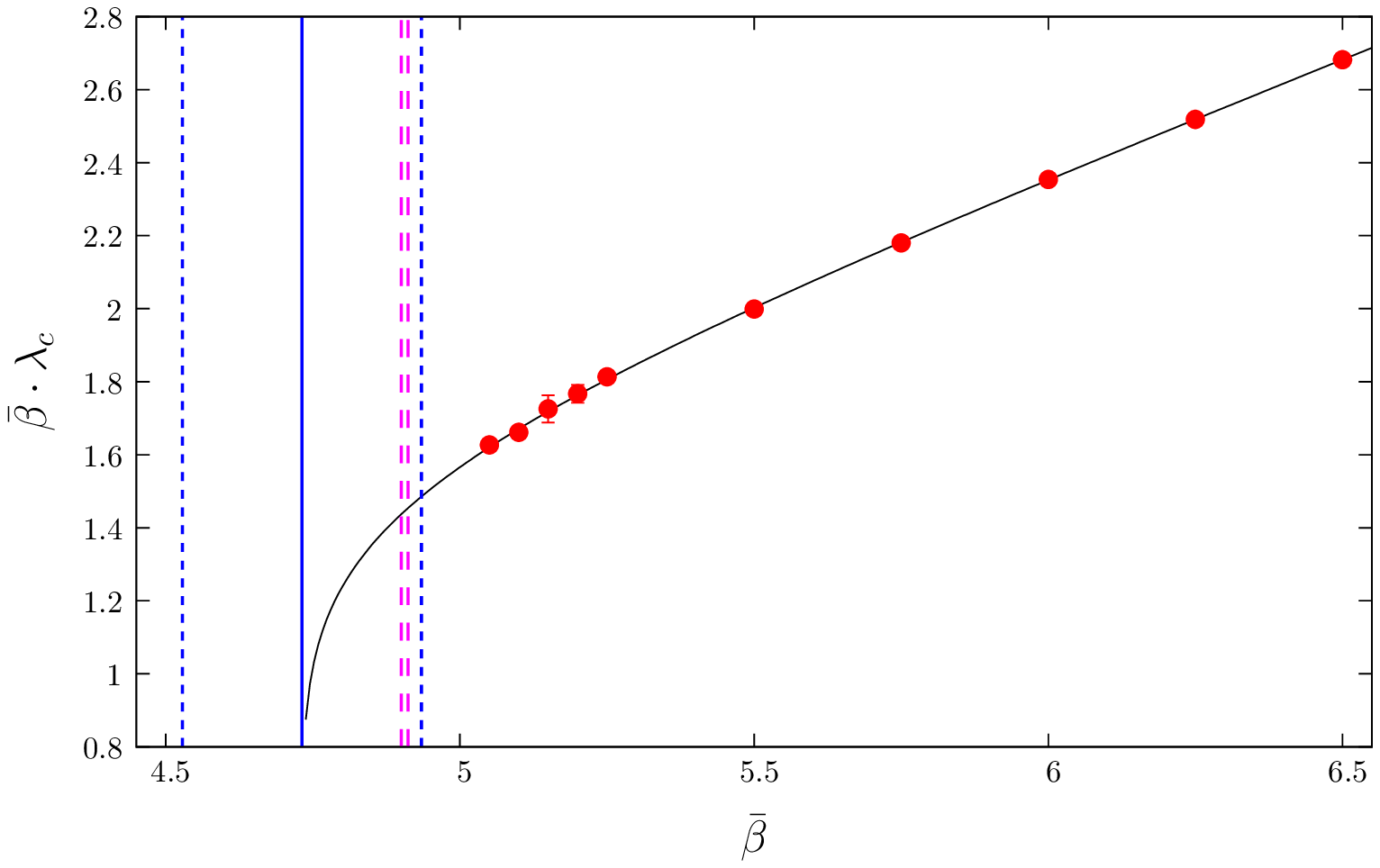}

\vspace{\floatsep}\vspace{\floatsep}
\includegraphics[width=0.75\textwidth]{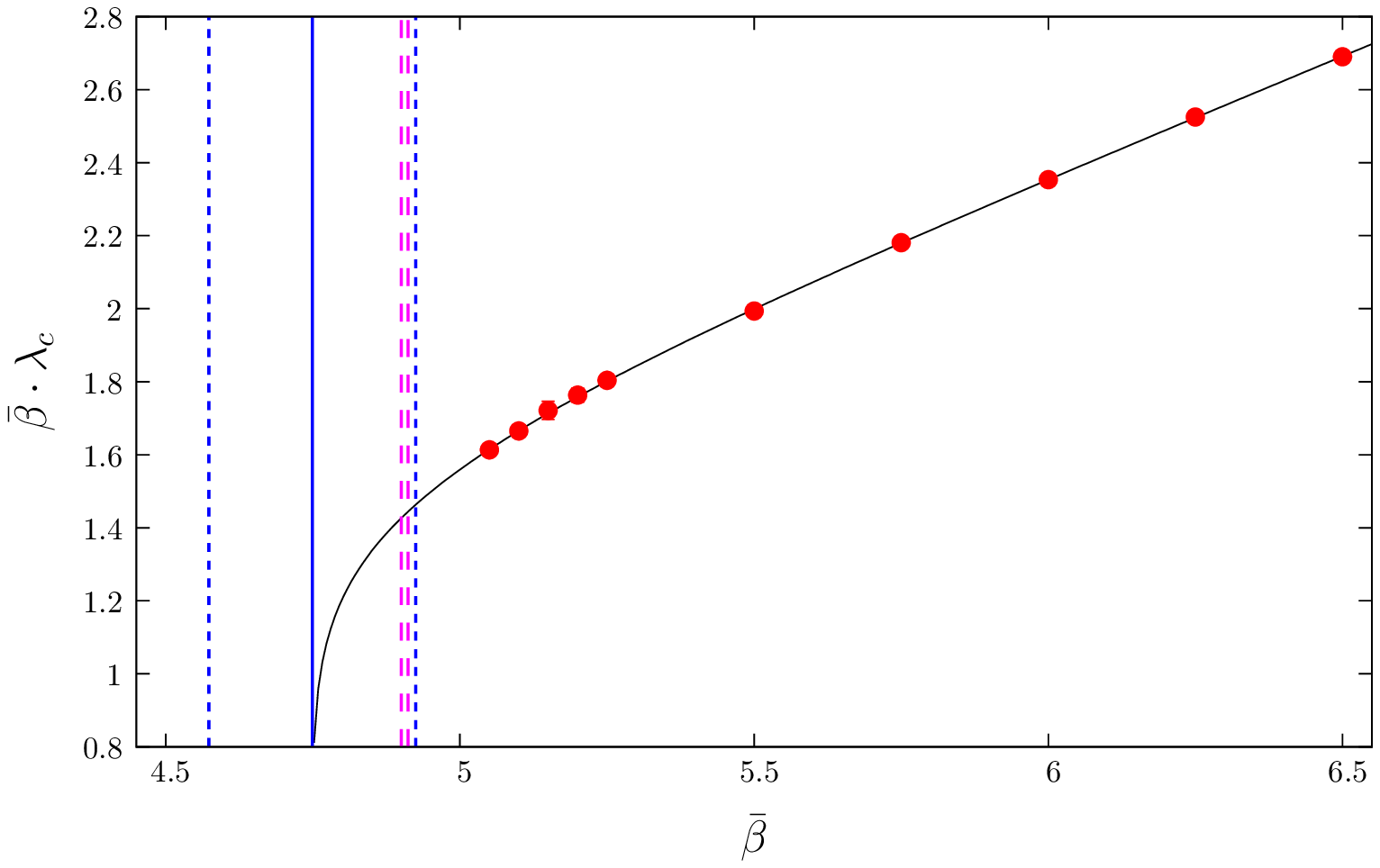}
  \caption{Mobility edge $\lambda_{c\,{\rm phys}}=\bred \cdot \lambda_c$ in
    physical units, as obtained fitting $I_\lambda$ (top) and $\la
    s^2\ra_\lambda$ (bottom). The function $\bred h(\bred)$ [see
    Eq.~\eqref{eq:lambdac_beta}] obtained fitting $\lambda_c$ is also
    shown (solid line). The positions of $\bred_0$ and corresponding
    error band (blue line) and the error band of $\bred_c$ (magenta
    line) are also shown. 
}   
  \label{fig:crit_lc_phys}
\end{figure}

The final results were obtained by collecting all the central values
obtained with the choice $(w,N_{s\,{\rm min}},\sigma)=(0.06,40,8)$. The
final error was obtained by adding in quadrature the statistical
error (as estimated by MINUIT\footnote{The symmetric parabolic error
  was used, since it never differed from the MINOS errors by more than
  10\%.}) and the three systematic errors, namely those related to the
width of the prior distribution, width of the fitting interval, and
minimal lattice size. In turn, these were estimated as the absolute
value of the difference between the final results and the central
values obtained with $(w,N_{s\,{\rm min}},\sigma)=(0.06,40,5)$,
$(0.055,40,8)$, and $(0.06,48,8)$, respectively. Among the various
fitting parameters, the most important ones are the mobility edge,
$\lambda_c$, and the values of the statistics at the critical point,
$I_c\equiv I_{\lambda_c}$ and  $\la s^2\ra_c \equiv \la
s^2\ra_{\lambda_c}$. These are shown in Figs.~\ref{fig:crit_I} and
\ref{fig:crit_s2}. The mobility edge in physical units,
$\lambda_{c\,{\rm phys}}(\bred)=\bred \lambda_c(\bred)$, obtained multiplying
$\lambda_c$ by the lattice coupling (assuming perfect scaling) is
shown in Fig.~\ref{fig:crit_lc_phys}. The values of $\lambda_c(\bred)$
obtained from the two spectral statistics agree with each other within 
errors. The values of the spectral statistics at the mobility edge
differ from those corresponding to Poisson or RMT statistics, and
depend on the lattice coupling (see Figs.~\ref{fig:crit_I} and
\ref{fig:crit_s2}). As functions of $\bred$ they are
rather flat at large $\bred$, while they seem to grow as one
approaches the critical temperature. This is different from what has
been observed in QCD and QCD-like models in 3+1 dimensions: there the
statistics at the mobility edge have always been found to be compatible
with the critical statistics of the three-dimensional unitary
Anderson model~\cite{Nishigaki:2013uya,Giordano:2016nuu}. 

\begin{figure}[t]
  \centering
  \includegraphics[width=0.49\textwidth]{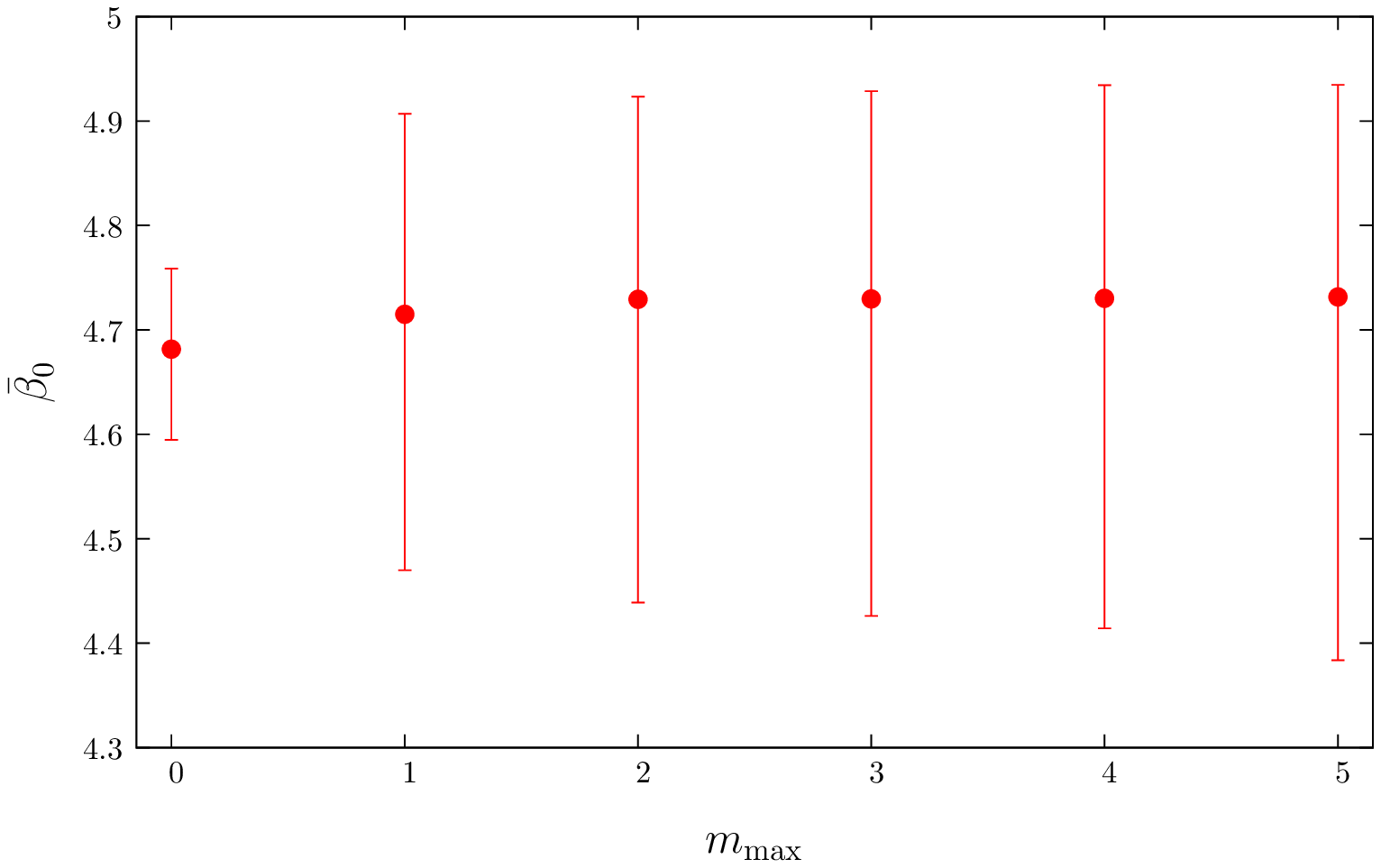}\hfil  \includegraphics[width=0.49\textwidth]{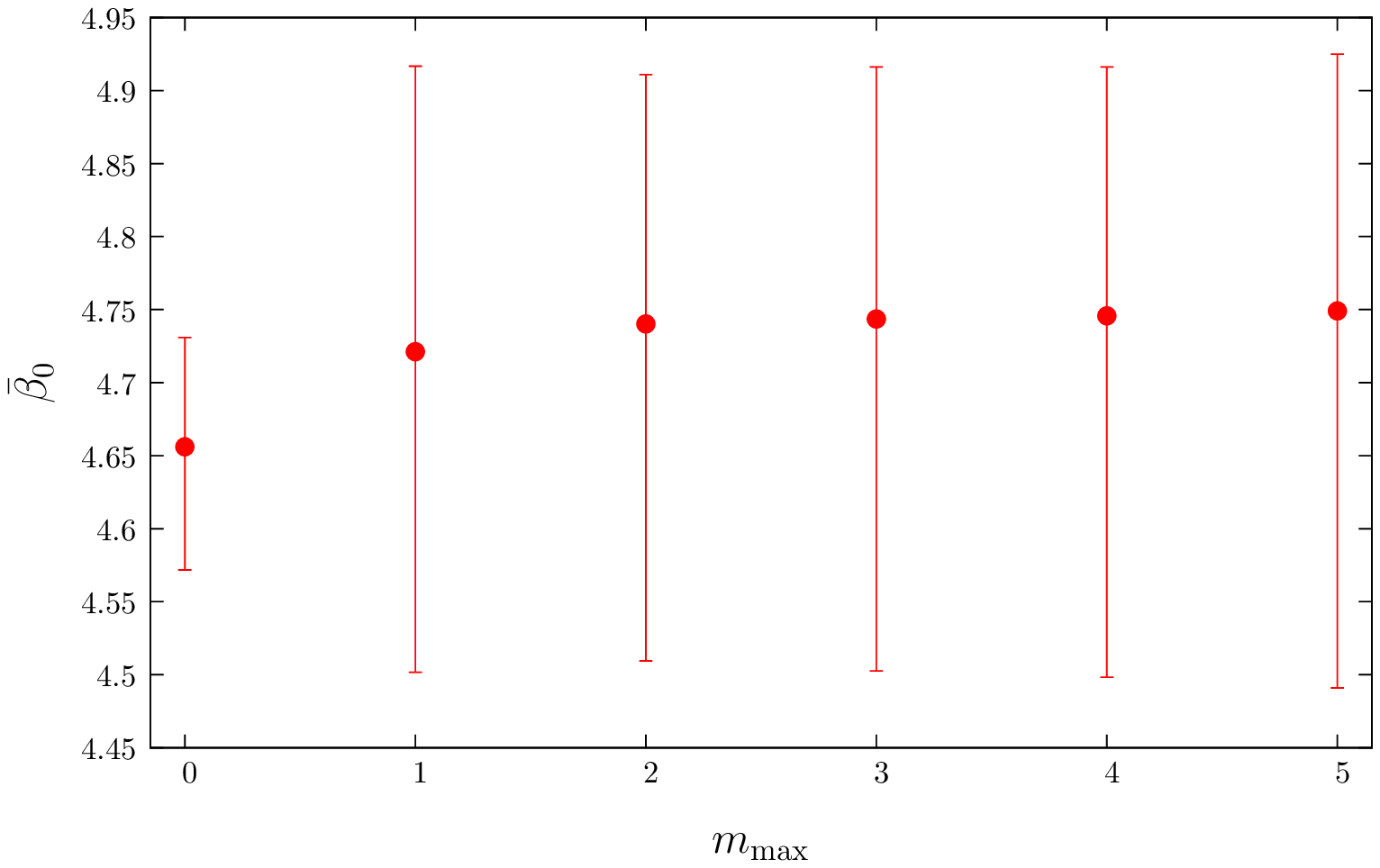} 
  \caption{Dependence of $\bred_0$ and $v$ on $m_{\rm max}$ in a
    constrained fit of $\lambda_c(\bred)$, obtained from
    $I_\lambda$ (left) and $\la s^2\ra_\lambda$ (right), with the
    function Eq.~\eqref{eq:lambdac_beta}.}  
  \label{fig:constrfit_beta0}
\end{figure}

Finally, I fitted the mobility edge in lattice units,
$\lambda_c(\bred)$, as obtained from the two spectral statistics, as a
function of $\bred$, trying to establish if it extrapolates to zero at
a coupling compatible with the critical one. In order to estimate
accurately the errors, I performed another constrained fit with a
function of the form 
\begin{equation}
  \label{eq:lambdac_beta}
  h(\bred) = u\,(\bred-\bred_0)^v \left(1 +
    \sum_{m=1}^{m_{\rm max}}h_m(\bred-\bred_0)^m\right)\,, 
\end{equation}
increasing $m_{\rm max}$ up to $m_{\rm max}=5$. While no prior was
assumed on $u$, $v$ and $\bred_0$, the parameters $h_m$ were
constrained to be small, centring their distributions at zero and
using widths $\sigma_1=\sigma_2=0.05$, $\sigma_3=0.005$,
$\sigma_4=0.001$, and $\sigma_5=0.0005$. These values were chosen so
that the fit would converge and errors would be estimated accurately
by MINOS. While the resulting errors on $h_m$ are bigger than the
central values, so that these coefficients are compatible with zero,
nevertheless their presence in the fit has a visible impact on the
errors of the important parameters. This can be seen in
Fig.~\ref{fig:constrfit_beta0}, where I show the dependence on $m_{\rm
  max}$ of $\bred_0$ and its error. The final results for $u$, $v$ and
$\bred_0$, determined from both the $I_\lambda$ and the $\la
s^2\ra_\lambda$ analyses, are reported in Table
\ref{tab:finalres}. The $\chi_{\rm aug}^2/n_{\rm data}$ is around
$0.2$ in both cases, a suspiciously small value that can however be
explained as follows.\footnote{It must be noted that for constrained
  fits the quantity to check is $\chi_{\rm aug}^2/n_{\rm data}$ with
  $n_{\rm data}$ the number of data points, and that while this
  quantity should be around 1, it is the convergence of the errors on
  the fit parameters that determines the quality of the
  fits~\cite{Lepage:2001ym}.} In Figs.~\ref{fig:crit_I} 
and  \ref{fig:crit_s2} one sees that the data points at
$\bred=5.15,5.20$ have a considerably larger error than the
other points. This is due to a rather large finite-volume effect:
values of $\lambda_c$ determined with $N_{s\,{\rm min}}=48$ 
show a visible jump upwards with respect to those determined with
$N_{s\,{\rm min}}=40$, something that does not happen for the other
values of $\bred$. This suggests that it is probably due to a
fit instability that would be cured by increasing the statistics, and
thus that the finite-volume error is actually overestimated. The
results obtained at $\bred=5.15,5.20$ with $N_{s\,{\rm min}}=40$ are
probably more reliable than those obtained with $N_{s\,{\rm min}}=48$,
and so data with $N_{s\,{\rm min}}=40$ have been used as central
values. Repeating the fit with Eq.~\eqref{eq:lambdac_beta} using these
data and only the statistical errors leads to similar results with a
$\chi_{\rm aug}^2/n_{\rm data}$ of about $0.7$--$0.8$.

\begin{table}[t]
  \centering
  \begin{tabular}{c|cc}
parameter & from $I_\lambda$ & from $\la s^2\ra_\lambda$ \\
\hline
    $u$            &  $0.377_{-0.017}^{+0.014}$ & $0.375_{-0.012}^{+0.013}$ \\
    $v$            &  $0.141_{-0.053}^{+0.12}$ & $0.135_{-0.049}^{+0.095}$ \\
$\bred_0$          &  $4.73_{-0.35}^{+0.15}$ & $4.75_{-0.26}^{+0.13}$
  \end{tabular}
  \caption{Results for $\bred_0$, $u$, and $v$ obtained with a
    constrained fit of $\lambda_c(\bred)$, obtained from
    $I_\lambda$ (left) and $\la s^2\ra_\lambda$ (right), with the
    function Eq.~\eqref{eq:lambdac_beta}, using $m_{\rm max}=5$.
}
  \label{tab:finalres}
\end{table}

In Figs.~\ref{fig:crit_I} and \ref{fig:crit_s2} I also show the fit to
the $\lambda_c$ data with Eq.~\eqref{eq:lambdac_beta}, marking the
critical point $\bred_0$ at which the mobility edge vanishes, and
the corresponding error band. The critical coupling $\bred_c$ for
deconfinement, as determined in Ref.~\cite{Liddle:2008kk}, is also
shown. The two values agree within one standard deviation, although 
they just do so. We could just be happy with that and conclude that
the mobility edge vanishes at the deconfinement transition, but it is
worth trying to explain why the difference is sizeable. According to
the sea/islands picture and the Dirac-Anderson approach, the source of 
disorder leading to localisation in the deconfined phase are the local
fluctuations of the Polyakov lines around the ordered value. One then 
expects the mobility edge to be sensitive to the average Polyakov
loop, since this provides a measure of how ordered the system
is. Finite-size effects affecting this quantity should then reflect  
themselves both on the mobility edge and on the pseudocritical
coupling in a finite volume. To try to quantify finite-size effects on
the latter quantity, I used the results of
Ref.~\cite{Liddle:2008kk}. There, the critical coupling in the
infinite-volume limit for the 2+1 dimensional SU(3) pure gauge theory 
is obtained via extrapolation from finite volumes using the formula  
\begin{equation}
  \label{eq:lt_extr}
  \f{\beta_c(\infty)-\beta_c(N_s)}{\beta_c(\infty)} =
h\left(\f{N_t}{N_s}\right)^{\f{6}{5}}\,,
\end{equation}
where $\beta_c(N_s)$ is the pseudocritical coupling defined as the
position of the peak of the Polyakov loop susceptibility,
$h$ is a fitting parameter, and the exponent is the one
appropriate for a second-order phase transition in the universality
class of the two-dimensional $q=3$ Potts model. Although not reported
explicitly, the value of $h$ for SU(3) can be estimated by using the
reported values for SU($N_c$) with $N_c=4,5,6,8$ and the approximate
behaviour $h(N_c)N_c^2 \simeq a+b/N_c^2$. Using the $N_t=4$ data, this
yields $h(3)\simeq 1.2$, and in turn a finite-volume pseudocritical
coupling  $\bred_c(72)\simeq 4.7$ for $N_s=72$, corresponding to the
largest volume used in this work. Although this value is remarkably
close to the results for $\bred_0$ in Table \ref{tab:finalres}, it
should not be taken too literally. The important point is that the
deviation from the infinite-volume limit is still about 4\%  for
$N_s=72$, which is just about the same deviation we find between
$\bred_0$ and the infinite-volume result for $\bred_c$. If anything,
this is reinforcing rather than weakening the claim that deconfinement
and localisation of the low modes happen together. In this respect, it 
might be worth mentioning that the exponent $v$ in Table
\ref{tab:finalres}, governing the approach to zero of the mobility 
edge, agrees (within the rather large errors) with the magnetisation
critical exponent governing the vanishing of the Polyakov loop
expectation value, which in turn has been found to agree with the
magnetisation critical exponent $\f{1}{9}$ of the two-dimensional
$q=3$ Potts model~\cite{Christensen:1991rx}. It would be interesting
to investigate this relation further, although this requires numerical
simulations close to the phase transition, where they are known to be
difficult.   

\begin{figure}[t]
  \centering
  \includegraphics[width=0.75\textwidth]{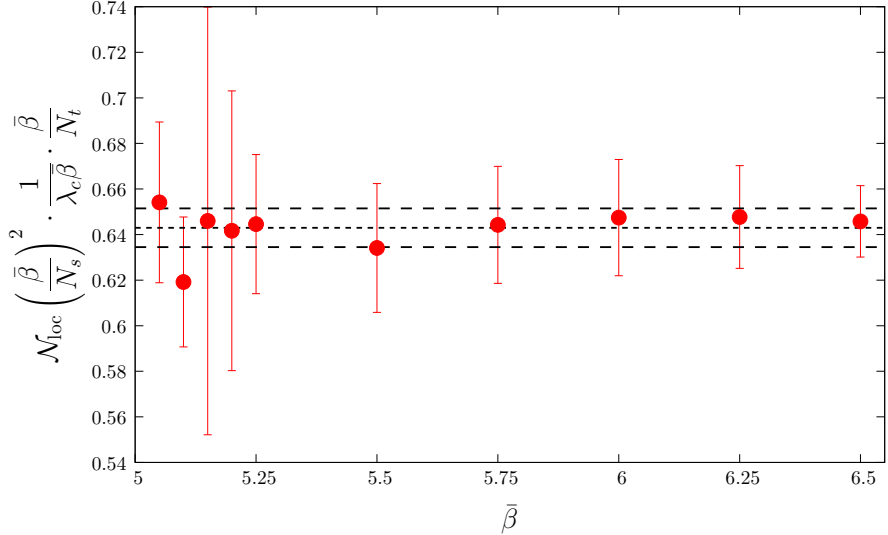}
  \caption{Density of localised modes times $T/\lambda_{c\, {\rm
        phys}}$ in physical units. A constant fit and the
    corresponding error band (dashed lines) are also shown.}  
  \label{fig:locmod}
\end{figure}

Once the mobility edge is known, the density of localised modes
in physical units can be computed as
\begin{equation}
  \label{eq:locdens}
  \f{{\cal N}_{\rm loc}}{V_{\rm phys}} = 
  \f{{\cal N}_{\rm loc}\,\bred^2}{V} =
  \left(\f{\bred}{N_s}\right)^2\int_0^{\lambda_c} d\lambda
  \,\rho(\lambda) \,,
\end{equation}
where both $\lambda$ and the spectral density $\rho(\lambda) = \la
\sum_n  \delta(\lambda-\lambda_n)\ra$ are in lattice units. In
Fig.~\ref{fig:locmod} I show this quantity multiplied by
$T/\lambda_{c\, {\rm phys}}=1/(N_t\lambda_c)$, i.e., 
\begin{equation}
  \label{eq:locdens2}
R \equiv  \f{{\cal N}_{\rm loc} T}{V_{\rm phys} \lambda_{c\,{\rm phys}}} =
  \left(\f{\bred}{N_s}\right)^2\f{1}{\lambda_c N_t}
  \int_0^{\lambda_c}d\lambda \,\rho(\lambda) \,.
\end{equation}
This quantity is independent of temperature within errors, and a
simple constant fit gives\footnote{Reinstating powers of $g^2$, one
  has $R=0.642(85)\, (g^2/2)^2$.} $R=0.642(85)$. This means that the
density of localised modes behaves like ${\cal N}_{\rm loc}/V_{\rm
  phys} \propto \lambda_{c\,{\rm phys}}/T \simeq (T -T_c)^v/ T$, with
$T_c=\bred_c/N_t$, and since $v\sim 0.1\div 0.2$ one has that it rises
steeply to a maximum, and then decreases quite fast with
temperature. This is different from what was found in QCD in 3+1 
dimensions~\cite{Kovacs:2012zq}: there, the density of localised modes
was seen to keep increasing, up to the highest available temperatures
of about $5\,T_c$, while here the decrease begins already at $1.1\div
1.2\,T_c$.   

\begin{figure}[t]
  \centering
  \includegraphics[width=0.75\textwidth]{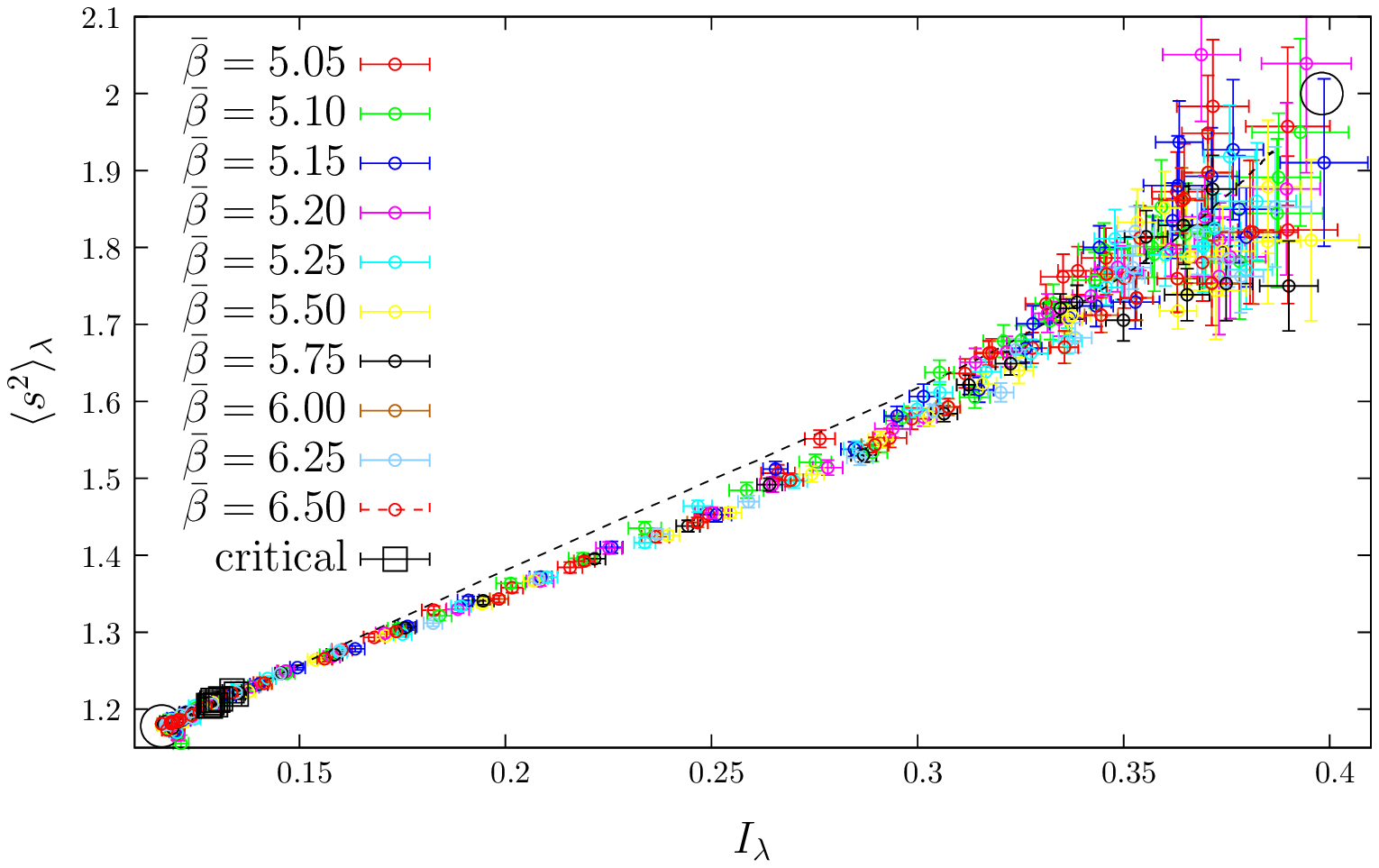}

\vspace{\floatsep}
\includegraphics[width=0.75\textwidth]{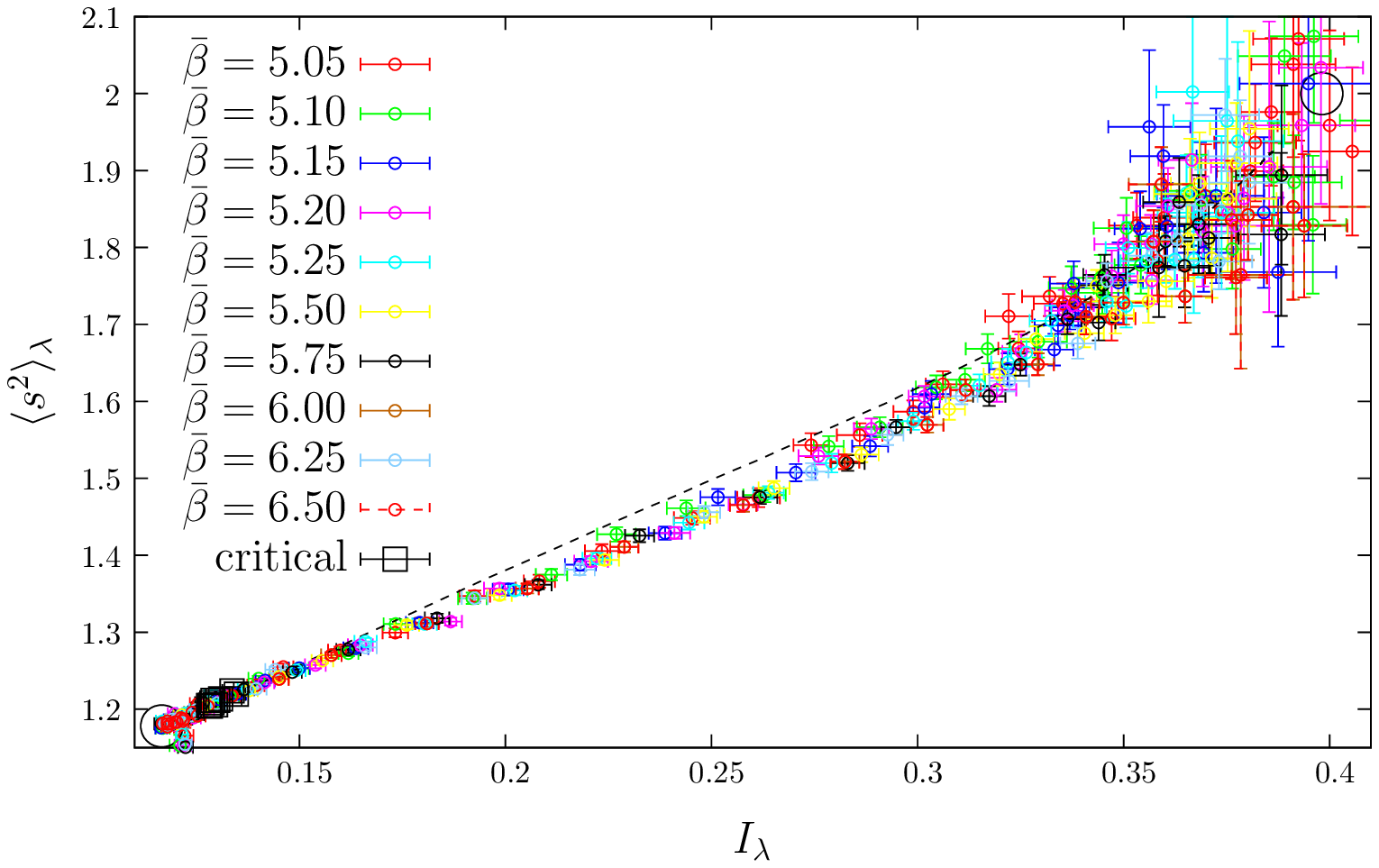}
  \caption{Shape analysis. Each data point corresponds to the pair
    $(I_\lambda,\la s^2\ra_\lambda)$ associated to the same point $\lambda$
    in the spectrum. The pairs corresponding to the mobility edges for
    the various couplings are shown as black squares. In the top panel
    $N_s=64$, while in the bottom panel $N_s=72$. Results for
    different lattice couplings and different volumes all lie
    approximately on the same curve. The points corresponding to RMT
    and Poisson statistics are denoted with circles. The curve
    corresponding to the sinh-kernel is shown with a dashed line.}  
  \label{fig:shape}
\end{figure}

To conclude the discussion of spectral statistics, I present an
independent test, usually referred to as {\it shape
  analysis}~\cite{varga1995shape}, for the one-parameter scaling 
hypothesis, Eq.~\eqref{eq:BKT_scaling}, exploited above to determine
the mobility edge by means of a finite-size scaling analysis. If
indeed a single quantity, namely the ratio $\log \xi/\log L$,
determines the statistical properties of the spectrum, then plotting
one spectral statistic against another should yield universal curves,
on which the data points coming from different volumes and lattice
couplings should all (approximately) lie, at least for sufficiently
large volumes. In Fig.~\ref{fig:shape} I show plots of $\la
s^2\ra_\lambda$ against $I_\lambda$ in the deconfined phase for
$N_s=64,\,72$. Only points for which $|\la s\ra_\lambda - 1|<0.05$ are
plotted. Data from different lattice couplings lie on a common curve
connecting the RMT point and the Poisson point. The same kind of
behaviour has been observed in 3+1-dimensional
QCD~\cite{Nishigaki:2013uya,Giordano:2014qna}. The dashed line
corresponds to the statistics determined by the so-called 
``sinh-kernel'',\footnote{I thank F.~Pittler for evaluating
  numerically the relevant quantities.} a one-parameter family of
two-level connected correlators that describes the  statistical
behaviour in the bulk of the spectrum for several different random
matrix models~\cite{Muttalib:1993zz,Mirlin:1996zz,Moshe:1994gc}. This
curve runs close to the numerical data obtained in 3+1 dimensional
QCD, intersecting them at the critical point~\cite{Nishigaki:2013uya}. 
In the present case, this curve describes well the numerical data
close to the RMT point, up to (and possibly even beyond) the critical
points found for the different values of the lattice coupling. This
suggests that the spectral statistics on the line of critical points
above the mobility edge belong to the family parameterised by the
sinh-kernel. Numerical errors are however still too large to make
conclusive statements.  

\begin{figure}[t]
  \centering
  \includegraphics[width=0.75\textwidth]{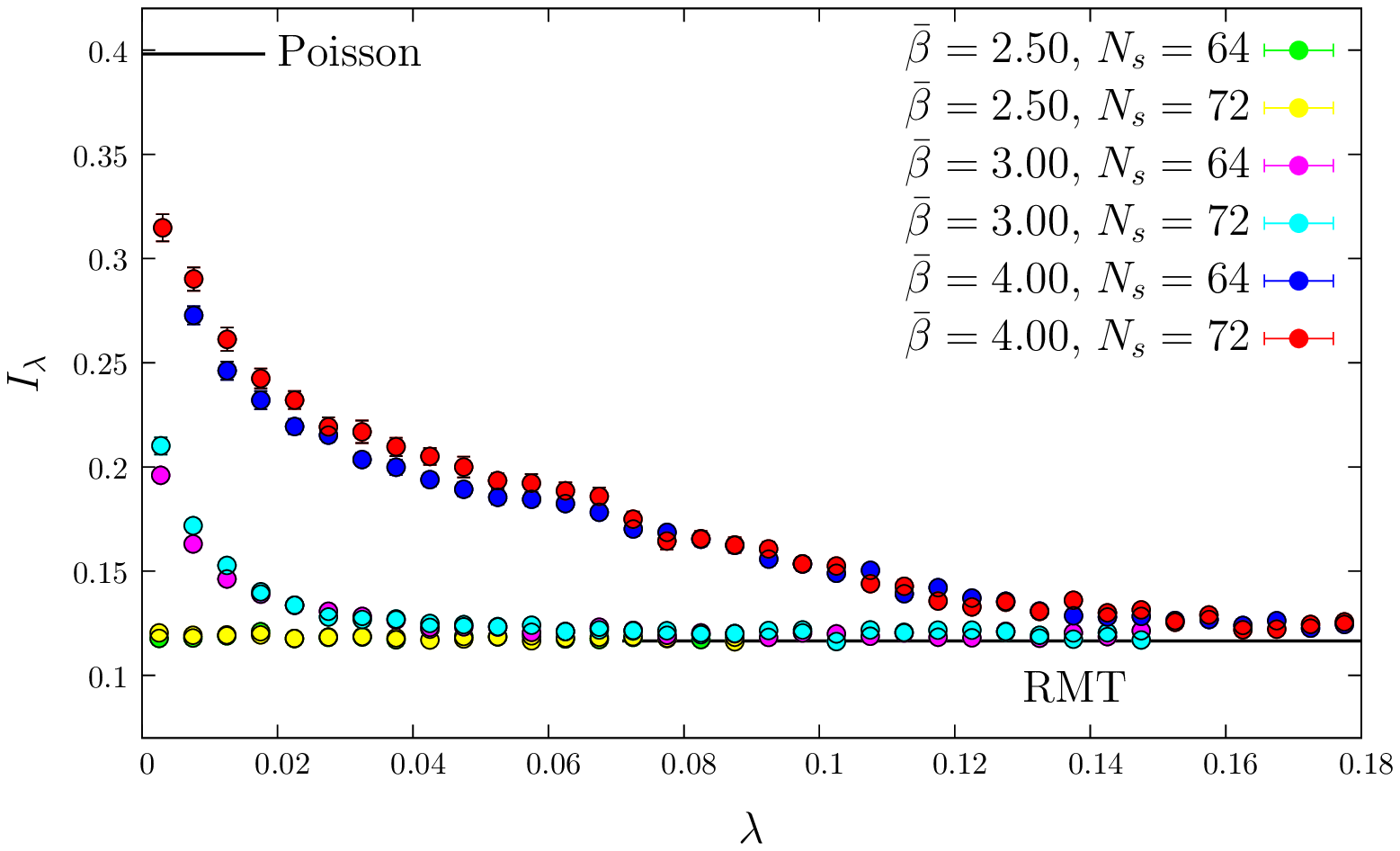}

\vspace{\floatsep}
  \includegraphics[width=0.75\textwidth]{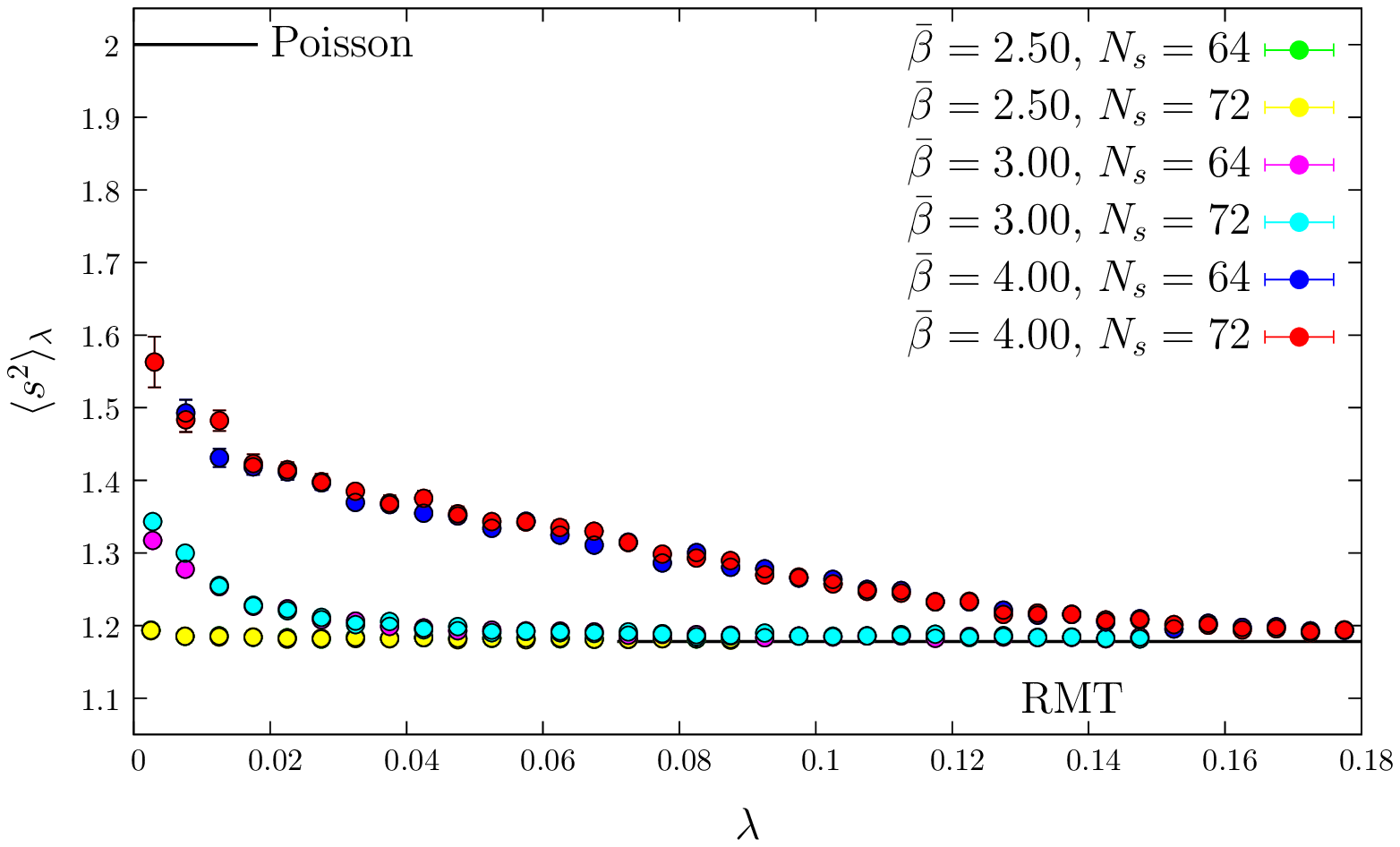}
  \caption{Spectral statistics $I_\lambda$ (top) and $\la
    s^2\ra_\lambda$ (bottom) in the confined phase, for the two
    largest lattice sizes.}  
  \label{fig:sp_low}
\end{figure}

\begin{figure}[t]
  \centering
   \includegraphics[width=0.75\textwidth]{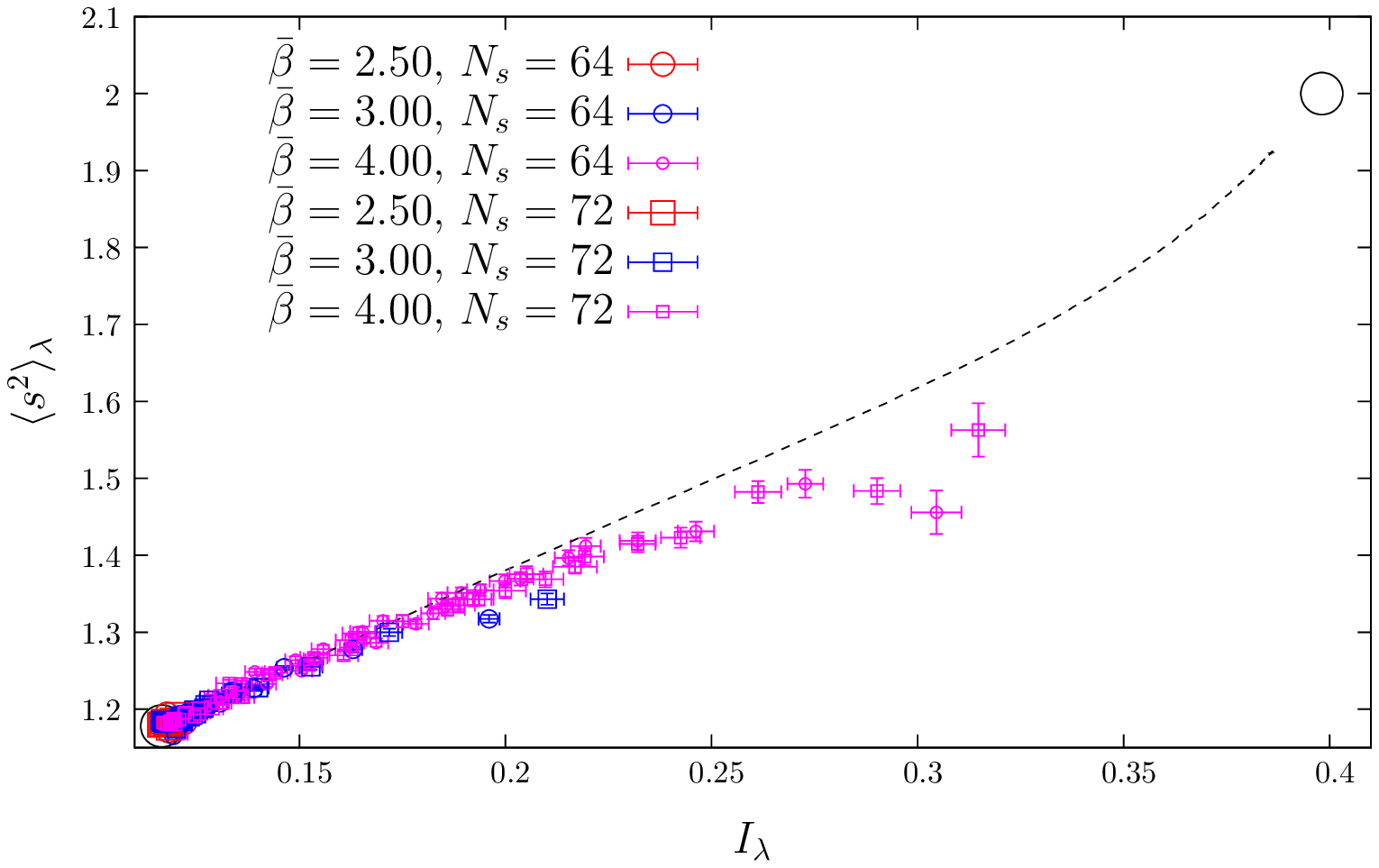}
  \caption{Shape analysis in the low-temperature phase. Only points with
    $|\la s\ra_\lambda - 1|<0.1$ are included.}  
  \label{fig:shape_low}
\end{figure}

After this long discussion of the high-temperature phase, a few words
about the low-temperature phase are in order. In Fig.~\ref{fig:sp_low}
I show the spectral statistics $I_\lambda$ and $\la s^2\ra_\lambda$ in
the confined phase. Analogously to what was found for the fractal 
dimension of the eigenmodes, the spectral statistics do not show a
uniform RMT-type behaviour, but they get closer to Poisson behaviour
in the low end of the spectrum, the more so as one gets closer to the
deconfinement transition. The transition in the spectrum however does
not seem to be a genuine phase transition: the distance from Poisson
behaviour remains large even for the largest volumes (as can be seen
in the shape-analysis plot in Fig.~\ref{fig:shape_low}), and attempts
at an unconstrained finite-size scaling analysis fail. It is likely
that in the infinite-volume limit the spectral statistics converge to
a non-trivial behaviour interpolating between Poisson and RMT: this
would reflect the non-trivial fractal dimension of the
eigenmodes.\footnote{Notice that what is being discussed here are bulk
  spectral statistics near the origin, and not microscopic spectral
  statistics, which require a different type of unfolding (see, e.g.,
  Ref.~\cite{Verbaarschot:2000dy}). For a comparison between the RMT
  predictions for the microscopic statistics and lattice data (at zero
  temperature) cf.\ Ref.~\cite{Bialas:2010hb}.}  
 
Combining the behaviour of the spectral statistics observed in the
confined phase with that observed at the mobility edge in the 
deconfined phase, one is led to conjecture the following development
with temperature. The spectral statistics at the low end of the
spectrum keeps approaching the Poisson statistics as the deconfinement
transition is approached from below, and reaches it at the critical
lattice coupling. At the same time, the crossover along the spectrum
turns into a true BKT-type, localisation/delocalisation phase 
transition, and a mobility edge appears at the origin. Moving further 
into the deconfined phase, the mobility edge moves up in the spectrum
and the statistics at the mobility edge moves towards RMT, apparently 
stabilising at an intermediate point between Poisson and RMT
for sufficiently large temperature.

\section{Conclusions and outlook}
\label{sec:concl}

The connection between deconfinement, chiral symmetry restoration and
localisation of the low Dirac modes has become increasingly evident
in recent years, both in
QCD~\cite{GarciaGarcia:2005vj,GarciaGarcia:2006gr,Kovacs:2009zj,   
  Kovacs:2012zq,Giordano:2013taa,Ujfalusi:2015nha,Cossu:2016scb,
Holicki:2018sms} and in QCD-like
theories~\cite{Kovacs:2010wx,Giordano:2016cjs,Giordano:2016nuu,
Giordano:2016vhx,Kovacs:2017uiz}. The theoretical arguments of
Refs.~\cite{Bruckmann:2011cc,Giordano:2015vla,Giordano:2016cjs}  
suggest that the driving force behind both chiral symmetry restoration 
and localisation is the ordering of the Polyakov lines causing 
deconfinement. This is further supported by numerical studies of
dedicated toy models~\cite{Giordano:2016cjs}. Given the generality of
these arguments, the connection between these three phenomena is
expected to be of rather general nature.  

A non-trivial test of this idea has been provided in the present
paper, devoted to the study of localisation of Dirac modes in 2+1
dimensional SU(3) pure gauge theory at finite temperature on the 
lattice. This model differs from QCD and from the other QCD-like
models mentioned above in several aspects: different dimensionality
(2+1 instead of 3+1), different critical behaviour at deconfinement
(second-order phase transition instead of crossover or
first order),\footnote{The only exception is the SU(2) case studied in
  Ref.~\cite{Kovacs:2010wx}, where however no detailed study of the
  transition region was made.} different expected type of
localisation/delocalisation transition (BKT instead of second
order). While the simultaneity of deconfinement and chiral restoration 
has long been known~\cite{Damgaard:1998yv}, no previous studies
existed about localisation of the Dirac modes. The numerical results
presented here, obtained with the staggered discretisation, indicate
that the lowest Dirac modes are delocalised (although with non-trivial
fractal dimension) in the confined phase, and localised in the
deconfined phase. A BKT-type Anderson transition is seen to take place
at a critical point (mobility edge) in the spectrum in the
high-temperature phase, in agreement with expectations based on
universality arguments and on known results for the two-dimensional
unitary Anderson model~\cite{xie1998kosterlitz}. Although this is
perfectly natural in the framework of the Dirac-Anderson
approach~\cite{Giordano:2016cjs}, it is by itself a rather nontrivial
finding, which provides nontrivial support to the related sea/islands
mechanism~\cite{Bruckmann:2011cc,Giordano:2015vla} for localisation in 
high-temperature gauge theories. Both the inverse of the typical size
of localised modes and the mobility edge extrapolate to zero at
temperatures compatible with the deconfinement temperature, 
as determined in Ref.~\cite{Liddle:2008kk}, indicating that the onset
of localisation coincides with deconfinement and chiral symmetry
restoration. This work thus provides further support to the idea of
deconfinement driving the system to a chirally restored phase with
localised low Dirac modes. 

There are other theories where it would be worth studying the relation
between deconfinement, chiral symmetry restoration and
localisation. The case of gauge theories in the presence of an
imaginary chemical potential is interesting both for its theoretical
aspects, and for the possible relevance to the study of hadronic
matter at finite density. Since the imaginary chemical potential
changes the effective boundary conditions affecting the Dirac
eigenmodes, according to the sea/islands picture and the
Dirac-Anderson approach it should control the density and the
localisation properties of the low modes. Preliminary
results~\cite{Giordano:2018iei} show that in SU(3) pure gauge theory 
in 2+1 dimensions an imaginary chemical potential leads to an increase
both in the spectral density near the origin and in the size of the
low modes, eventually leading to chiral-symmetry breaking and
delocalisation of the low modes for sufficiently large imaginary
chemical potential, in agreement with expectations. 

Another interesting case is that of U(1) pure gauge theory in 2+1
dimensions, for various reasons. The Abelian nature of the gauge group
and the different nature (BKT) of the finite-temperature deconfining
transition provide a setup qualitatively different from any other
investigated so far. Moreover, in contrast with the case of SU($N_c$)
gauge groups, in the high-temperature phase there is no infinite
barrier in the thermodynamic limit separating Polyakov loop sectors
that differ by the phase of the spatially-averaged Polyakov loop. 
Since it is the phases of the Polyakov lines that provide the
effective boundary conditions for the Dirac modes, in turn determining
their localisation properties, one expects the coexistence of
localised and delocalised modes at the low end of the spectrum. Also
in this case the available preliminary results~\cite{Giordano:2018iei}
confirm the expectations. 

An important consequence of localisation of the low modes is that it
prevents a spontaneously broken continuous symmetry from generating
Goldstone bosons~\cite{McKane:1980fs,Golterman:2003qe}. If the low
Dirac modes become localised at deconfinement then the Goldstone
mechanism does not apply anymore, independently of the spectral
density near the origin, and Goldstone bosons disappear from the
spectrum. From this point of view, the connection between
deconfinement and localisation is possibly even more important than
that between localisation and chiral symmetry breaking. A setting in
which these issues can be investigated is the SU(3) gauge theory with
adjoint fermions in 3+1 dimensions, which has long been known to
display separate deconfining and chirally-restoring phase
transitions~\cite{Karsch:1998qj}. Preliminary (unpublished) results 
show that this happens also in 2+1 dimensions with gauge group SU(2)
in the quenched approximation, and that while the spectral density
remains finite above the deconfinement transition, the near-zero modes
become localised.

The results of these paper increase the confusion about the role
played by topology in making the low Dirac modes localised. In
Refs.~\cite{GarciaGarcia:2005vj,GarciaGarcia:2006gr} the localised 
modes were understood as coming from the localised zero modes
supported by instantons at finite temperature. However, the authors of
Ref.~\cite{Kovacs:2010wx} claimed that the estimated density of
topological objects was an order of magnitude smaller than that of
localised modes (in quenched SU(2) configurations). More recently, in
Ref.~\cite{Kovacs:2017uiz} (see also Ref.~\cite{Kovacs:2019txb}) the
density of topological near-zero modes in the pure gauge SU(3) theory
was shown to account only up to 60\% of the localised modes at the
deconfinement transition, a fraction rapidly falling with temperature,
and which overestimates the corresponding result in the presence of
dynamical fermions. The results of Ref.~\cite{Cossu:2016scb} (for QCD
with domain-wall fermions) show that localised modes prefer locations
with larger action density and topological charge density, which may
be related to the positions of $L$-$\bar{L}$ monopole-instanton
pairs. On the other hand, in 2+1 dimensions there is no topological
charge for SU(3) fields, since the homotopy group $\pi_2({\rm SU}(3))$
is trivial, and nonetheless localised modes appear in the deconfined
phase. Further studies are needed to clarify this issue, which, given
the close relation between localisation and deconfinement, might even
help in understanding the role played by topology in the deconfinement 
transition. 

In conclusion, further studies of localisation in gauge
theories may help in better understanding the relation between
confinement and chiral symmetry breaking.

\section*{Acknowledgements}

I wish to thank M.~Caselle, J.~Greensite, D.~N\'ogr\'adi, F.~Pittler and
L.~von Smekal for useful discussions, and T.~G.~Kov\'acs both for
discussions and for a careful reading of the manuscript. This work was
partly supported by grants OTKA-K-113034 and NKFIH-KKP126769.

\bibliographystyle{JHEP}
\bibliography{references_loc}

\end{document}